\documentclass{article}

     \usepackage[final,nonatbib]{neurips_2024}

\usepackage[utf8]{inputenc} % allow utf-8 input
\usepackage[T1]{fontenc}    % use 8-bit T1 fonts
\usepackage{hyperref}       % hyperlinks
\usepackage{url}            % simple URL typesetting
\usepackage{booktabs}       % professional-quality tables
\usepackage{amsfonts}       % blackboard math symbols
\usepackage{nicefrac}       % compact symbols for 1/2, etc.
\usepackage{microtype}      % microtypography
\usepackage{xcolor}         % colors
\usepackage{multirow}
\usepackage{amsmath}
\usepackage{graphicx}
\usepackage{chemformula}
\usepackage{braket}
\usepackage[style=numeric-comp,sorting=none,maxbibnames=9999]{biblatex}
\usepackage{algorithmic}
\usepackage[ruled,vlined]{algorithm2e}
\usepackage{amsmath}
\usepackage{paralist}
\title{Integrating Graph Neural Networks %and First Principles Theory %Quantum Mechanics Density Functional Theory 
and Many-Body Expansion Theory 
for %Full-Dimensional 
Potential Energy Surfaces}

\author{%
\textbf{Siqi Chen}$^{\textbf{1},\ast}$ \quad \textbf{Zhiqiang Wang}$^{\textbf{1,2},\ast}$ \quad \textbf{Xianqi Deng}$^{\textbf{1,3},\ast}$ \quad Yili Shen$^{\textbf{1,4},\ast}$ \quad \textbf{Cheng-Wei Ju}$^\textbf{1,5}$\\
\textbf{Jun Yi}$^\textbf{1,6}$ \quad \textbf{Lin Xiong}$^\textbf{1}$ \quad \textbf{Guo Ling}$^\textbf{1}$ \quad \textbf{Dieaa Alhmoud}$^\textbf{1}$ \qquad \textbf{Hui Guan}$^{\textbf{1},\dagger}$ \qquad \textbf{Zhou Lin}$^{\textbf{1},\dagger}$\\
$^\textbf{1}$University of Massachusetts Amherst \qquad 
$^\textbf{2}$Florida Atlantic University \qquad%, Boca Raton, FL 33431\\
$^\textbf{3}$University at Albany \\ %, Albany, NY 12222 \\\
$^\textbf{4}$University of Notre Dame \qquad %, Notre Dame, IN 46556\\
$^\textbf{5}$The University of Chicago \qquad %, Chicago, IL 60637\\
$^\textbf{6}$Wake Forest University\\%, Winston-Salem NC 27109 \\
$^\ast$These authors contribute equally to this study. \\ 
$^\dagger$Corresponding authors: \texttt{\{huiguan,zhoulin\}@umass.edu}\\
}

\begin{document}

\maketitle

\begin{abstract}
%\zhou{Siqi: Do you have a paid account to invite Xianqi and other authors?}
Rational design of next-generation functional materials %pushes the scientific and technological frontiers and 
relied on %computational and 
quantitative predictions of their 
%The advancement of functional materials like metal-organic frameworks (MOFs), organic semiconductors (OSCs), and RNA sequences, pivotal for next-generation technologies, relying on unraveling their complex 
%quantum mechanical 
electronic structures beyond single building blocks. 
First-principles quantum mechanical (QM) modeling %model, density functional theory %s 
%has emerged 、
%emerges as a robust and accurate %, and efficient 
%computational tool for %evaluating 
%electronic structures of a small molecule, % due to a balanced accuracy and efficiency, 
%but 
became %expensive or 
infeasible as the size of a material grew beyond hundreds of atoms. %even at the ground state. %these properties but is computationally intensive. To address this, machine learning (ML) algorithms, such as neural networks (NNs), are being used to bypass the bottleneck. However, most NN models fall short in accurate molecular representation and lack the latest graph neural network (GNN) incorporations. 
In this study, we developed a new computational tool integrating fragment-based graph neural networks (FBGNN) into the fragment-based many-body expansion (MBE) theory, % and the subgraph-based graph neural networks (FBGNNs), 
referred to as FBGNN-MBE, and demonstrated its capacity to reproduce full-dimensional potential energy surfaces (FD-PES) %of a large material 
for hierarchic chemical systems with manageable accuracy, complexity, and interpretability. % in a divide-and-conquer manner. 
%\zhou{Siqi: We probably need a name for our method. Do you have any idea?}
%combining deep learning with quantum chemistry to predict electronic structures and provide design principles for materials like MOFs and OSCs. 
In particular, we divided the entire system into basic building blocks (fragments), evaluated their single-fragment energies using a first-principles QM model %, such as second-order M{\o}ller-Plesset perturbation theory (MP2) or density functional theory (DFT), 
and attacked many-fragment interactions using the structure--property relationships trained by FBGNNs. %subgraph-based graph neural networks (\textcolor{red}{FBGNNs}). 
%advanced GNN models into many-body expansion (MBE) theory to manage computational complexity, evaluating high-dimensional energy surfaces for key materials. 
%By breaking materials into fragments, we used DFT for single-fragment energies and GNN-trained models for multi-fragment interactions.
%We trained and tested the first version of FBGNN-MBE on big clusters of water (\ch{H2O}), phenol (\ch{C6H5OH}), and 1:1 water--phenol mixture.
%Trained and tested on the clusters of water, phenol, and water--phenol mixture, our FBGNN-MBE %method 
%Our approach achieved chemical accuracy compared with pure first-principle counterparts in determining the full-dimensional potential energies of these clusters, with mean absolute errors (MAE) under 1.0 kcal/mol \textcolor{red}{for XXXXXXXXX}.
%\zhou{Siqi: It is not clear what system you are talking about for these MAEs. We want to reflect what we have done.}
%Meanwhile, our approach significantly reduced %ing computational overhead while maintaining accuracy. 
%the computational cost by \textcolor{red}{XX.XX, XX.XX, and XX.XX} orders of magnitude for these three systems.
%In addition, the implementation of subgraphs into GNNs notably enhanced their performance and interpretability for systems exhibiting chemical hierarchy like ours. 
Our development of FBGNN-MBE demonstrated the potential of a new framework integrating deep learning models into fragment-based QM methods, and 
%demonstrating the potential of integrating GNNs with DFT through MBE theory for predictive modeling of complex materials. The development of GNN-MBE 
marked a significant step towards computationally aided design of large functional materials.
%\zhou{Siqi: I think our abstract is still too long.}
\end{abstract}

\section{Introduction}
%\zhou{When you make comments, please use a different color so I can be aware of it.}
% \paragraph{Computational Material Discovery}

Discovery of complex materials that exhibited exceptional quantum mechanical (QM) properties and function beyond single monomers %building blocks and fully 
and equilibrium structures, such as %ionic liquids~\cite{https://doi.org/10.1002/adma.202313023}, 
metal--organic frameworks (MOF)~\cite{yao2020metal}, 
organic semiconductors (OSC)~\cite{doi:10.1021/acs.accounts.3c00750}, and %even 
branched deoxyribonucleic acids (DNA)~\cite{doi:10.1021/acs.chemrev.0c00294}, 
was crucial %in the breakthrough 
in emergent scientific and technological areas, % of %modern science and technology, 
such as carbon neutrality~\cite{doi:10.1021/accountsmr.2c00084}, renewable energy~\cite{doi:10.1126/science.adq3799}, and next-generation optoelectronics~\cite{https://doi.org/10.1002/advs.202003834}. % and medical sciences~\cite{tang2020materials}.
Computational chemistry eliminated %extremely time-consuming and 
%resource-intensive 
expensive trial-and-error experiments and %s in wet-lab experimentation and 
explored the vast chemical space. % for design and optimization. 
In the present study, we aimed to accomplish a computational design
for these complex materials based on their aggregate and dynamic QM properties,  
%chemistry %discovery 
which required a rapid and rigorous evaluation of their full-dimensional potential energy surfaces (FD-PES) on the fly.
%In this context, f
This job cannot be done by first-principles QM models like %, such as 
second-order M{\o}ller--Plesset perturbation theory (MP2)~\cite{PhysRev.46.618} %,head1988mp2} 
or density functional theory (DFT)~\cite{PhysRev.136.B864,PhysRev.140.A1133} %,parr1994density},
%offer rigorous solution to predict electronic structures of complex chemical systems 
%without experimental inputs, and their popularity keeps increasing along with the advancement in newer methods~\cite{10.3389/fchem.2021.705762} and faster supercomputers~\cite{doi:https://doi.org/10.1002/9781394197705.ch18}.
%Successful examples include the constrained DFT (CDFT) for the charge-transfer states~\cite{doi:10.1021/cr200148b}, 
%optimally-tuned range-separated hybrid (OT-RSH) density functionals for the %accurate fundamental and optical 
%band gaps of conjugated organic molecules and charge-separated complexes~\cite{doi:10.1021/ct2009363}, 
%orbital optimized density functional theory (OO-DFT) for the %precise descriptions of 
%charge-transfer states, doubly 
%multiply excited states%, and core-level excited states
%~\cite{doi:10.1021/acs.jpclett.1c00744}, and energy decomposition analysis (EDA) for the %quantitative analyses of 
%energy components in chemical bonds and 
%non-bonding interactions~\cite{https://doi.org/10.1002/wcms.1345}.
%advancements in exchange-correlation functionals~\cite{PhysRev.136.B864,PhysRev.140.A1133}, successfully capturing interactions in complex systems. For example, DFT has been instrumental in identifying the interaction between metal monolayers and oxide supports, and in understanding stability based on band structures~\cite{o2018interaction}. 
%Given the power of MP2 and DFT, 
due to the
%However, MP2 and DFT are 
%still 
%computationally 
prohibitive costs for %complex and interacting chemical systems like 
large systems %functional materials 
because their computational complexity scaled as the fifth and third power of the number of % cubically with the number of 
basis functions~\cite{doi:https://doi.org/10.1002/9781119019572.ch14,C5CP00437C}. 

%,nagai2018neural}.
Motivated by this problem, many fragment-based ``divide-and-conquer'' methods were
developed %under the philosophy of ``divide-and-conquer'' 
to accelerate typical QM approaches %like MP2 or DFT %reduce the computational complexity of quantum mechanical models 
%and enable them for %describing 
%large systems 
while maintaining the accuracy~\cite{doi:10.1021/acs.accounts.6b00356,10.1063/1.5126216}. %,D0CP01095B}. 
%Among cost-effective approximation methods for quantum mechanical modeling, 
%\zhou{Siqi: I found you sometimes do not write concisely. AI may not include useful information.}
Among all these theories, 
%Among all fragment-based quantum mechanical approaches, 
many-body expansion (MBE) stood %stands 
out due to its straightforward implementation and rapid convergence for many-body %(many-fragment) 
interactions% and rapid convergence in the higher-order terms
~\cite{doi:10.1021/ct700223r,doi:10.1021/ar500119q,%10.1063/5.0057752,
heindel2022many}.
%by employing a "divide-and-conquer" fragment-based approach, which reduces computational costs while maintaining accuracy. 
MBE partitioned a %large and 
complex %chemical 
system into %small and 
manageable fragments (bodies) and expanded the total electronic energy or %electronic energy or 
other relevant properties into a series of one-body (1B) and many-body ($n$B) terms %interactions 
with progressively diminishing contributions. % of these interactions %expensive but 
%decreasingly significant contributions. %~\cite{varandas1987double}. This structured approach facilitates the calculation of a system's energy and other essential properties, 
This hierarchical treatment not only streamlined a calculation with a reduced computational complexity but also enabled a deeper analysis of the electronic structure landscape and the %, highlighting 
intricate %intra- and inter
many-fragment interactions, % that are crucial for discovering new materials.
%In addition to reducing the computational complexity, this treatment allows us to analyze the details of electronic structure landscape and %allowing for a detailed analysis of a molecule's energy landscape and capturing 
%capture the nuanced %intra- and inter
%many-fragment interactions, 
both of which were critical properties for computational material discovery. 

%\zhou{Siqi: We need to summarize the conclusions and the works from two or more important groups about MBE. Would you like to have a try based on the papers I gave you here?}
%
The Herbert group and the Xantheas group 
made prominent and complementary contributions in recent methodology of MBE for %and have created systematic applications of MBE in 
both static and dynamic behaviors of condensed-phase systems.
%In recent years, significant contributions to MBE methods have come from two prominent research groups, each advancing the field in different but complementary directions. 
%In some studies, 
Herbert and coworkers developed the generalized many-body expansion (GMBE) framework to
%has developed a framework that generalizes traditional MBE approaches to 
handle %more complex systems, including those 
%complex 
systems with ill-defined or overlapping fragments like %, such as %. Their introduction of the Generalized Many-Body Expansion (GMBE) allows for the consideration of overlapping fragments, which is particularly important when dealing with systems where fragment boundaries are not well-defined, such as in 
%water clusters or 
fluoride-water complexes. 
%Additionally, the Herbert group has 
They also introduced energy-screened MBE %approaches to %techniques to improve computational efficiency, 
with enhanced efficiency and intact accuracy
%reduce the computational cost without impacting the accuracy 
by selectively including only sizable many-body contributions %those 
%higher-order terms that significantly contribute to the 
in the total electronic energy %This approach enhances the method's robustness and accuracy, making it more feasible to apply MBE to larger, more complex molecular systems 
%For example, Herbert and coworkers \textcolor{red}{present a unified framework for fragment-based methods in electronic structure theory, introducing the Generalized Many-Body Expansion (GMBE) to handle overlapping fragments in systems  like water clusters and fluoride-water complexes, and energy-screened MBE for enhanced cost efficiency, offering improved accuracy and robustness over traditional methods.} 
~\cite{10.1063/1.4742816,doi:10.1021/ct300985h,
doi:10.1021/jz401368u,10.1063/1.4885846,doi:10.1021/ar500119q,
10.1063/1.4947087,
doi:10.1021/acs.jctc.5b00955,10.1063/1.4986110,doi:10.1021/acs.jctc.9b01095,
10.1063/5.0174293}.
%Simultaneously, 
Xantheas and coworkers leveraged MBE for %full-dimensional 
potential energy surfaces (%FD-
PES) %and molecular dynamics (MD) simulations 
and %, emphasizing the method's ability to capture non-additive effects in complex systems. Their work 
demonstrated that MBE can provide a more quantitative understanding of molecular properties than simpler pairwise-additive models. % can offer. The Xantheas group has 
They further %developed innovative protocols for incorporating 
incorporated MBE into molecular dynamics (MD) simulations to involve %allow for the consideration of %, which allow for accurate modeling of systems with significant 
subtle QM phenomena for electrons and nuclei%nuclear quantum effects and other subtle QM phenomena
~\cite{heindel2020many,doi:10.1021/acs.jctc.0c01309,doi:10.1021/acs.jctc.1c00780,10.1063/5.0095335,heindel2022many,D1CP00409C,10.1063/5.0095739,doi:10.1021/acs.jctc.3c00575,D2CP03241D,doi:10.1021/acs.jpclett.2c03822,10.1063/5.0094598,PhysRevC.107.044004}.
%. This has broadened the applicability of MBE beyond static systems, enabling researchers to explore both ground-state properties and dynamic behaviors in condensed-phase environments 
%Xantheas and coworkers implemented MBE methods and \textcolor{red}{demonstrate the effectiveness of MBE methods in understanding molecular properties and capturing non-additive effects in water clusters. They further introduce new MBE protocols for molecular dynamics simulations to accurately model complex systems, including light nuclear interactions.}~\cite{heindel2020many,doi:10.1021/acs.jctc.0c01309,doi:10.1021/acs.jctc.1c00780,10.1063/5.0095335,heindel2022many,D1CP00409C,10.1063/5.0095739,doi:10.1021/acs.jctc.3c00575,D2CP03241D,doi:10.1021/acs.jpclett.2c03822,10.1063/5.0094598,PhysRevC.107.044004}.
Despite these advances, applying QM-based MBE to functional materials with more sophisticated structures and more intense interactions than water clusters %or ice crystals beyond 
remains a challenge %. This is primarily 
due to %high computational demands of 
the large numbers of %of QM %first-principles 
%evaluations %of a 
%large number of 
%for all %possible %$n$-mers in 
$n$-fragment interactions for high $n$'s. %, %which require fully converged self-consistent field (SCF) calculations. As these systems grow and complexity, the associated computational costs become prohibitive, limiting the 
The integration of 
%To address these challenges, 
%machine learning (ML) algorithms, especially %such as 
neural networks (NNs) offered a revolutionary approach to accelerate QM methods like MBE% first-principles calculations including MBE for QM properties %and model quantum mechanical properties
~\cite{%behler2007generalized,gilmer2017neural,
schutt2017quantum,keith2021combining,https://doi.org/10.1002/wcms.1645}.
In particular, Parkhill and coworkers merged NN into MBE (NN-MBE) %, and their approach 
and demonstrated its strong predictive power for the FD-PES %full-dimensional potential energy surface (FD-PES) for 
of methanol (\ch{CH3OH}) clusters with %, reaching small 
mean absolute errors (MAEs) of 9.79 and 12.55 kcal/mol for two-body (2B) and three-body (3B) energies compared to MP2 but %, and %, and within the chemical accuracy} and \textcolor{red}{
a reduced computational cost by six orders of magnitude%a factor of $10^6$ compared to MP2%traditional ab initio methods like MP2-MBE
%} %potential in accurately predicting energy surfaces for large molecular clusters with reduced computation time, as demonstrated in their proof-of-concept using methanol 
~\cite{yao2017many}. 
%and facilitate MBE calculations in specific~\cite{yao2017many}. 
%\zhou{Siqi: Not all examples here are designed for quantum mechanics. I saved non-QM examples for later.}
However, %some 
intrinsic problems of traditional NNs %associated with traditional NNs, such as the lack of 
in terms of the missing physical information% like QM laws%from quantum mechanics
~\cite{schutt2017quantum,doi:10.1126/sciadv.1603015},
%the lack of 
the limited transferability %\xianqi{we have two "r" here: The limited transferability} 
and interpretability~\cite{gilmer2017neural,C7SC04934J}, 
%the in
and inability to handle graph-structured data~\cite{kipf2016semi,NIPS2017_5dd9db5e} %, along with other numerical issues~\cite{bengio2012practical}, 
%make them suffer 
%the difficulty 
%in capturing QM %quantum mechanical 
%properties, 
%NNs have demonstrated remarkable efficiency in modeling quantum mechanical properties, reducing computational overhead significantly. 
%However, challenges such as difficult convergence, frequent collapse, vanishing gradients, and hyperparameter sensitivity~\cite{bengio2012practical} 
compromised their capacity in %complex landscape of 
%NN-based quantum chemistry 
QM modeling~\cite{behler2007generalized,doi:10.1126/science.aag2302}. % and %. These issues 
%highlight an urgent need for method development. % of alternative ML approaches. %further development that can navigate these obstacles while preserving the integrity and accuracy of the models.

Instead, the development of graph neural networks (GNNs) %exhibited 
experienced exceptional success %capacity in modeling 
in chemical systems %molecules and materials 
because their node--edge structures naturally aligned with %represent three-dimensional 
three-dimensional atom--bond structures and encoded mechanical information about chemical bonds and intermolecular interactions
%They have emerged as a new family of machine learning tools in chemical systems because their node–edge structures naturally represent the three-dimensional (3D) atom–bond structures.
%As a result, they exhibit
%were selected for this study due to their 
%This character leads to exceptional ability in capturing 
%model complex molecular structures and capture 
%intricate interactions between building blocks and incorporating crucial chemical features
~\cite{kipf2016semi,NIPS2017_5dd9db5e,chen2019graph,dai2021graph,dai2021graph-corr}. 
%Also, just like plain NNs, GNNs are also able to process large data sets.
Outstanding examples included %This includes 
SchNet~\cite{schutt2018schnet}, GeoMol~\cite{ganea2021geomol}, FP-GNN (fingerprints-GNN)~\cite{cai2022fp}, and dyMEAN (dynamic multi-channel equivariant graph network)~\cite{kong2023end} which incorporated %, with incorporation of 
complex geometric %or spatial 
information in the graph representation, %known for its 
%effective graph-based atomistic representation %and continuous-filter convolutional layers~\cite{schutt2018schnet}, %effectiveness in modeling atomistic systems; 
PhysNet~\cite{unke2019physnet}, DimeNet/DimeNet++ (directional message passing NN)~\cite{gasteiger2020directional,gasteiger2020fast}, %,zhu2023fastdimenet++}, 
E(n) EGNN (equivariant GNN)~\cite{satorras2021n}, SEGNN (steerable E(3) equivariant GNN)~\cite{brandstetter2021geometric}, and ViSNet (vector-scalar interactive GNN)~\cite{wang2024enhancing} which integrated %, %known for its 
%with integration 
directional message passing framework and physical principles, %which excels in energy and force predictions; 
and ml-QM-GNN (QM-augmented GNN)~\cite{stuyver2022quantum}, MD-GNN (mechanism-data-driven graph neural network)~\cite{chen2023md}, MP-GNN (multiphysical GNN)~\cite{li2022multiphysical}, and SS-GNN (simple-structured graph neural network)~\cite{zhang2023ss} which implemented %, with implementation of 
quantitative mechanical and electronic properties. %and enhanced prediction of chemical reactivity%, known for significantly enhancing the prediction of chemical reactivity through the integration of quantum mechanical descriptors such as partial charges and NMR shielding constants and machine learning
%FP-GNN (fingerprints and GNN), with integrated molecular fingerprints as descriptors and captured structural and environmental information%, known for its innovative integration of molecular graphs and fingerprints, significantly enhances the prediction of molecular properties, utilizing molecular graphs to capture structural information and local atomic environments while employing molecular fingerprints to provide complementary chemical features
%~\cite{cai2022fp}. 
%MP-GNN (multiphysical GNN), with incorporated multiscale interactions embedded into element-specific graph representations%known for systematically representing various molecular interactions and efficiently processing complex data by utilizing a unique architecture that incorporates both scale-specific and element-specific graphs
%~\cite{li2022multiphysical}. 
%dyMEAN (dynamic Multi-channel Equivariant Graph Network), known for a dynamic Multi-channel Equivariant Graph Network to process the structural information~\cite{kong2023end}. 
%SS-GNN (Simple-Structured Graph Neural Network), known for offering efficient and accurate drug-target binding affinity predictions through a simplified architecture, minimizing complexity while maximizing performance~\cite{zhang2023ss}. 
%MD-GNN (mechanism-data-driven graph neural network), known for enhancing molecular properties prediction by effectively integrating molecular graph structures with additional electronic and structural information~\cite{chen2023md}.
%}
Most of these GNN models %achieve 
demonstrated enhanced performance in molecular representation learning %and chemical property prediction, %generating properties from structures, 
but 
%However, most of these GNN models 
they treated all atoms %or fragments 
on equal footing without considering the chemical hierarchy, which impacted their descriptive and predictive capacity for complex systems with many building blocks.
%\zhou{Siqi: we missed the discussion about subgraph-based GNN here, and the works of Zhang or maybe someone else should be mentioned here. Can you also find a few papers and summarize them here? There should be many in biophysics.}

%Instead, a group of 
State-of-the-art GNN models with subgraph of fragment-based frameworks, %that demonstrate frameworks of subgraphs, 
such as SubGNN (subgraph NN)~\cite{alsentzer2020subgraph}, FragGraph~\cite{doi:10.1021/acs.jpca.1c06152}, subGE (subgraph embedding)~\cite{chen2023subge}, MXMNet (multiplex molecular GNN)~\cite{zhang2020molecular}, and PAMNet (physics aware multiplex GNN)~\cite{zhang2023universal},
all represented %place material 
building blocks like molecules or monomers into subgraphs or local graphs, and captured interatomic, intermolecular and interfragment interactions using local and global message-passing architectures. %, respectively. 
%\siqi{Subgraph Neural Networks (SubGNN) represent a framework for subgraph classification, effectively capturing complex local relationships within subgraphs by utilizing a message-passing architecture, making it particularly useful in domains like chemistry for tasks such as molecular property prediction and drug discovery~\cite{alsentzer2020subgraph}.} 
%Compared to the original version of GNN, Subgraph-based GNN (\textcolor{red}{subGNN}) models, designed to model \textcolor{red}{model local and non-local interactions within complex molecular structures}~~\cite{zhang2020molecular,zhang2023universal}, became 
Such an analogy %alignment 
between hierarchic graph structures and hierarchic chemical systems rendered these models outstanding %choices as the 
%backbones %approaches 
methods for studying complex systems. % across the %molecules and materials across the diverse compositional 
%diverse chemical space. % because their subgraph architecture can align with the non-local interactions in the many-fragment systems.
%Several renowned neural network architectures have been pivotal in computational chemistry. 
In particular, MXMNet %(molecular mechanics-driven graph neural network)~
%~\cite{zhang2020molecular} 
and PAMNet %(physics aware multiplex graph neural network)~
%~\cite{zhang2023universal} 
developed by Xie and coworkers %\textcolor{red}{have
significantly advanced the representation learning of hierarchic %molecular 
systems by integrating molecular mechanics and multiplex graph representations and proved successful in reproducing the molecular properties from the QM9 data set~\cite{QM9}, the protein--ligand binding affinities from the PDBBind data set~\cite{doi:10.1021/jm048957q}, and the three-dimensional (3D) structures of ribonucleic acids (RNA)~\cite{doi:10.1126/science.abe5650}.

%FragGraph developed by Raghavachari and coworkers, \textcolor{red}{integrated fragmentation, error cancellation, and deep learning to enhance accuracy in predicting molecular properties such as thermochemical properties.}~\cite{doi:10.1021/acs.jpca.1c06152} \siqi{Qian and coworkers present subGE, a subgraph embedding framework that improves molecular property prediction. Their approach leverages subgraph representations, incorporates reinforcement learning for dimension reduction, and utilizes a mutual information mechanism to enhance predictive accuracy~\cite{chen2023subge}.}

%GNNs are particularly effective in generating embeddings of polycrystalline microstructures, incorporating physical features that are crucial in materials science~\cite{dai2021graph}.  
%Unlike conventional machine learning models, GNNs accurately capture non-local interactions within many-fragment systems~\cite{zhang2020molecular}. %These capabilities make GNNs an ideal choice for modeling materials across diverse compositional spaces while accurately representing complex, non-local interactions. 

%\zhou{The following citations are not used yet.}\cite{gassner1998representation,hansen2013assessment,doi:10.1021/ct600253j,dahlke2006assessment}

%\zhou{I copied the following sentence from results.}
%These models are at the forefront of molecular representation learning, selected for their proven efficacy and diverse approaches to chemical prediction tasks.

In the present study, we developed a novel computational model named FBGNN-MBE (fragment-based graph neural network driven many-body expansion) to address all problems mentioned above. %, %subGNN-MBE, 
%which married 
Our ultimate goal was to accomplish a rapid, precise, transferable, and interpretable scheme to evaluate FD-PES for any functional materials with many building blocks and important dynamic properties.  
%reduced computational complexity associated with traditional QM and QM-MBE methods without sacrificing the accuracy~\cite{jiang2021could,hamilton2020graph}.
Our major contributions include:
\begin{compactitem} 
  \item We established %the backbone idea of %The basic idea of 
  FBGNN-MBE %married 
%which 
to integrate the divide-and-conquer strategy of the MBE formalism with the sophisticated modeling capacity of FBGNN.
  \item We attacked the total ground state electronic energy using MBE and expected an extension to excited state energies and other properties. %, and truncated the expansion at the 3B terms. 
  \item We evaluated 1B energies using %first-principles 
  MP2 or DFT %calculations, and 
  and generated 2B and 3B energies based on 3D atomistic geometries and structure--property relationships trained by MXMNet %~\cite{zhang2020molecular} or 
  or PAMNet. %~\cite{zhang2023universal}. 
  \item We provided a proof-of-concept for FBGNN-MBE using three benchmark systems with weak to moderate many-fragment interactions. %, including water (\ch{H2O}) clusters, phenol (\ch{C6H5OH}) clusters, and 1:1 water--phenol (\ch{H2O:C6H5OH}) clusters. 
  \item We arrived at chemical accuracy ($<0.3$ kcal/mol) for 2B and 3B energies across all systems and outperformed other MBE models using conventional GNNs. 
  \item We interpreted the outstanding performance of FBGNN-MBE through the natures and strengths of many-fragment interactions in benchmark systems. 
  \item We designed application systems to evaluate FBGNN-MBE in reproducing experimental measurable properties based on FD-PES.
  \item We confirmed the potential of FBGNN-MBE as a revolutionary protocol for computational material discovery.
\end{compactitem} 

%MXMNet~\cite{zhang2020molecular} or PAMNet~\cite{zhang2023universal} in specific, into the original formalism of MBE. 
%Here we used MXMNet~\cite{zhang2020molecular} and PAMNet~\cite{zhang2023universal} as the backbones of our GNNs because 

%Graph Convolutional Networks (GCNs), a GNN variant, particularly have drawn significant attention because of their efficiency in processing large datasets and their superior ability to encode molecular intricacies\cite{kipf2016semi}. 
%This work explores subGNN-MBE's potential to revolutionize material discovery.
%This project pioneers a method that integrates GNNs with DFT through MBE theory, presents an innovative pathway to bypass the computational challenges inherent in quantum mechanics,  aiming to significantly enhance the computational exploration of functional materials. 

%\hui{Problems to address: 1) The paragraphs are all too long. Consider cut the paragraphs shorter or split one paragraphs to mulitples. 2) The part that describe our approach is too short.
%Typical intro flow: 
%what is the problem we want to solve and why the problem is important? what are the existing approaches and why they cannot solve the problem? what is our proposed solution? what is the basic idea and distinctive features of our solution? what are the contributions? It is important to use bullet points to highlight the contributions. 
%}

\section{Methods}

%In the present section, we will 
%We herein provided brief descriptions of the MBE formalism and %the two FBGNN approaches, the 
%the FBGNN-MBE method. %, %our subGNN-MBE models that integrate MXMNet and PAMNet into MBE, and %innovative subGNN-MBE model that integrates these concepts for improved scalability and accuracy, and 
%and the computational details.

\subsection{Many-Body Expansion Theory}
\begin{figure}[!ht]
\centering
\includegraphics[width=0.65\textwidth]{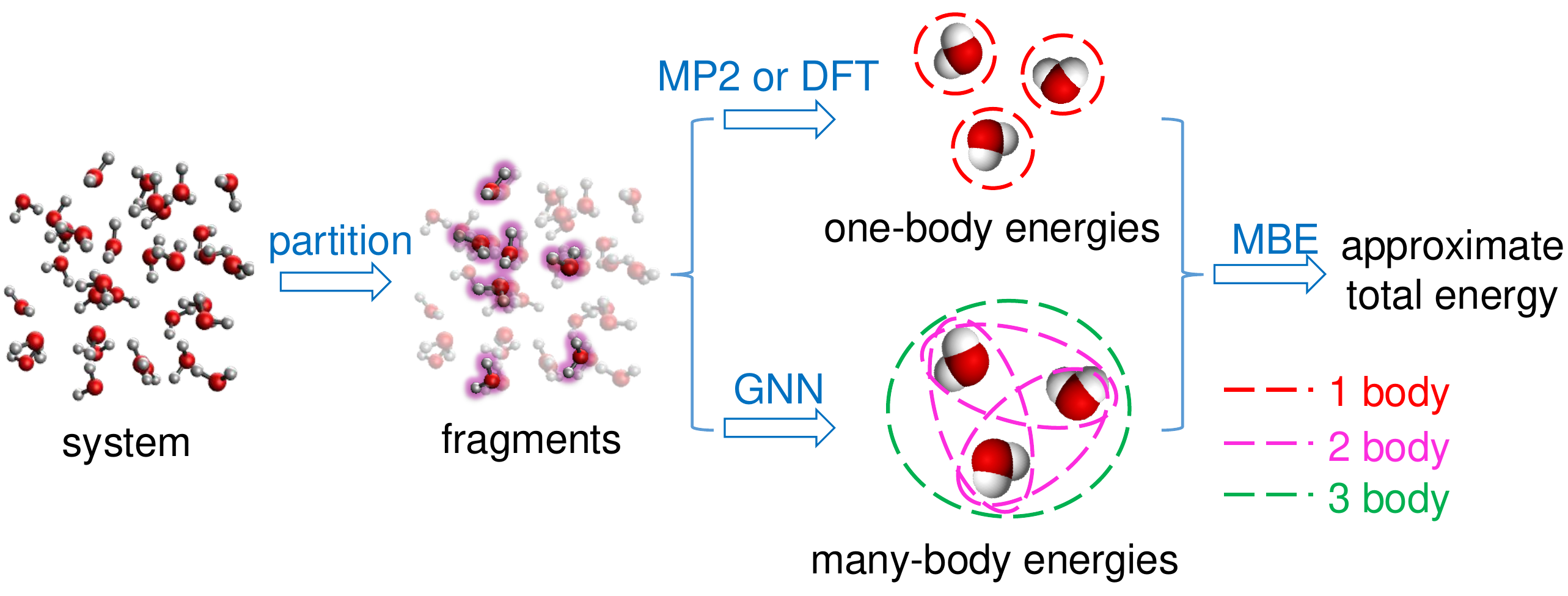}
\caption{Schematic strategy of our FBGNN-MBE approach integrating FBGNNs into the MBE theory, using a water cluster as an illustrative example.} % \zhou{Siqi: I revised this figure and uploaded the original pptx file. I think the PDF version has a higher resolution.}}
\label{fig:1}
\end{figure}

%Central to our computational methodology is the application of MBE, 
MBE %is a powerful fragment-based method %As the backbone of our method development, MBE is a powerful 
%tool 
%that 
decomposed any %an extensive or intensive 
aggregate %physical 
property, such as the total ground state energy ($E$), into %,  physical quantity which systematically calculates the energy 
contributions from individual fragments (1B) and many-fragment ($n$B) interactions. %their interactions between each other.
Truncated at a vanishing high-order term, %the implementation of 
MBE facilitated an efficient approximation of the property in question. %of a system's total energy.  
Beyond reducing the computational cost, MBE also enabled the capture of the natures and strengths of many-fragment interactions in a chemical system.
The generic %version of 
MBE theory partitioned a system %of interest 
into $N$ fragments and expands %and truncates the total energy to the 
$E$ until the a higher order tuncation (3 here)~\cite{%varandas1987double,gassner1998representation,10.1063/1.4742816,doi:10.1021/ct600253%j,10.1063/1.3121323,10.1063/1.4742816,doi:10.1021/ct400488x,
heindel2022many}:
\begin{equation}
E = \sum_{i}^{N}E^\text{1B}_{i} + \sum_{i<j}^{N}{E^\text{2B}_{ij}} + \sum_{i<j<k}^{N}{E^\text{3B}_{ijk}} %+ \ldots + \sum_{i<j<k<\ldots}^{N}{E^{M\text{B}}_{ijk\ldots}}, 
\label{eq:mbe}
\end{equation}
%Here 
$E^\text{1B}_{i}$ represented a 1B energy for %, evaluated using 
the isolated $i^\text{th}$ %single 
fragment. 
$E^\text{2B}_{ij}$ represented a 2B energy, capturing the interactions between the $i^\text{th}$ and $j^\text{th}$ fragments.
$E^\text{3B}_{ijk}$ represented a 3B energy, providing more complicated interactions among the $i^\text{th}$, $j^\text{th}$, and  $k^\text{th}$ fragments. As such 
\begin{align}
E^\text{1B}_{i} & = E_{i} \\
E^\text{2B}_{ij} & = E_{ij} - E^\text{1B}_{i} - E^\text{1B}_{j} \label{eq:2b}\\
E^\text{3B}_{ijk} & = E_{ijk} - E^\text{2B}_{ij} - E^\text{2B}_{ik} - E^\text{2B}_{jk} - E^\text{1B}_{i} - E^\text{1B}_{j} - E^\text{1B}_{k} \label{eq:3b} %\\
%\ldots & = \ldots
\end{align}
%As the expansion goes to a higher order, it 
Although higher-order terms captured %tends to capture more complex but more subtle 
more complex interactions, % within the system.
%In the present study we 
we %decided to 
%truncated at the 3B 
neglected %any terms 
beyond 3B energies due to vanishing contributions, tolerable errors, and exponentially-growing sizes. % due to practical considerable, including the controllable number of subsystems ($n$-mers) and the weaker non-covalent many-fragment interactions in the benchmark systems.
%to account for more complex correlations within the system, and
%In our version of 
We discussed our fragmentation strategies in Section \ref{sec:strategy}.
In FBGNN-MBE, we utilized a hybrid strategy by calculating 1B energies using MP2 or DFT but leveraging %the power of 
FBGNNs for 2B and 3B energies %, we utilize MBE to calculate the 2B and 3B energies for our dataset. We integrate DFT calculations for one-fragment properties and  leverage the predictive power of GNNs for many-fragment terms (Figure 1), thus avoiding computational constraints. This hybrid approach mitigates the need for extensive self-consistent field (SCF) calculations 
(Figure \ref{fig:1}).

\subsection{Fragment-Based Graph Neural Networks}

%{\color{red} 
%Our method fragments systems based on chemical properties, preserving functional units and bonds with bond orders above two; in this study, we focused on molecular aggregates, treating each molecule as a fragment to capture intermolecular interactions in $n$-body energies ($n \geq 2$). For future polymer studies, each monomer will be treated as a fragment. Our fragmentation strategy a chemical system was borrowed from earlier fragment-based studies like first-principles MBEs, such as  - Adaptive Systematic Molecular Fragmentation (ASMF), Generalized Many-Body Expansion (GMBE), and Energy-Screened Many-Body Expansion. These studies enhance the efficiency and accuracy of calculating molecular properties by decomposing complex systems into overlapping fragments, with ASMF allowing for systematic exploration of fragment contributions to molecular properties, GMBE ensuring unique accounting of interactions and Energy-Screened MBE selectively computing only significant interactions.}

\paragraph{General Architecture of GNNs}
%\zhou{Siqi: I also changed the subsubsection into a paragraph so that it can save some space.}%A GNN typically consists of a sequence of convolution layers.
%Each layer is a learnable non-linear function to transform from the input feature description to the output intermediate representation, or from the input intermediate representation to the output node embedding. 
%It then uses a graph pooling operation to reduce all node embedding vectors into a single graph embedding vector, so that it can reduce the size of the graphs, specifically the number of nodes, but keeping critical structural and features. 
%It finally uses both node embedding and graph embedding for downstream tasks including predicting two- and 3B energies in the present study (Figure \ref{fig:1.5})~\cite{kipf2016semi,NIPS2017_5dd9db5e,hamilton2020graph}
A GNN typically employed multiple convolution layers to transform input features into node embeddings, followed by graph pooling to generate a graph %-level 
representation, which can then be used for various downstream tasks such as predicting 2B and 3B energies~\cite{kipf2016semi,NIPS2017_5dd9db5e,hamilton2020graph}.
%add to Siqi: 
In %the context of 
chemical applications, the input graph was structured with features, including nodes representing individual atoms and edges capturing pairwise interatomic %, interatomic %pairwise relationships between these atoms, typically based on 
interactions as functions of atomistic structures. %of charges and distances. %bond lengths or other distance metrics. 
Each node was initialized with features that described its atomic properties, such as atomic number, charge, electronegativity, and even orbitals, %other relevant descriptors, 
while each edge reflected an interatomic distance or a bond length. %chemical bond. %These node and edge features serve as inputs to the GNN, which iteratively 
These features iteratively updated the node embeddings by aggregating information from neighboring nodes and effectively extract %captures effectively capturing the 
the local environment each atom resided in. %(I am not very sure about the specific information, please help me to check them.)
%\zhou{Siqi: We should discuss how much is needed for this paragraph. The discussion seems to be too general here. Please ask for advice from Prof. Guan.} 

% \begin{figure}[!ht]
% \centering
% \includegraphics[width=0.6\textwidth]{Figure1.5-GNN.pdf}
% \caption{Schematic design of a generic GNN with multiple convolutional layers.} %\zhou{Siqi: I think we can replace upper panel of the original Figure 2 with this one. Your version using phenol as an example may confuse computer scientists.}}
% \label{fig:1.5}
% \end{figure}

\begin{figure}[!ht]
\centering
\includegraphics[width=0.8\textwidth]{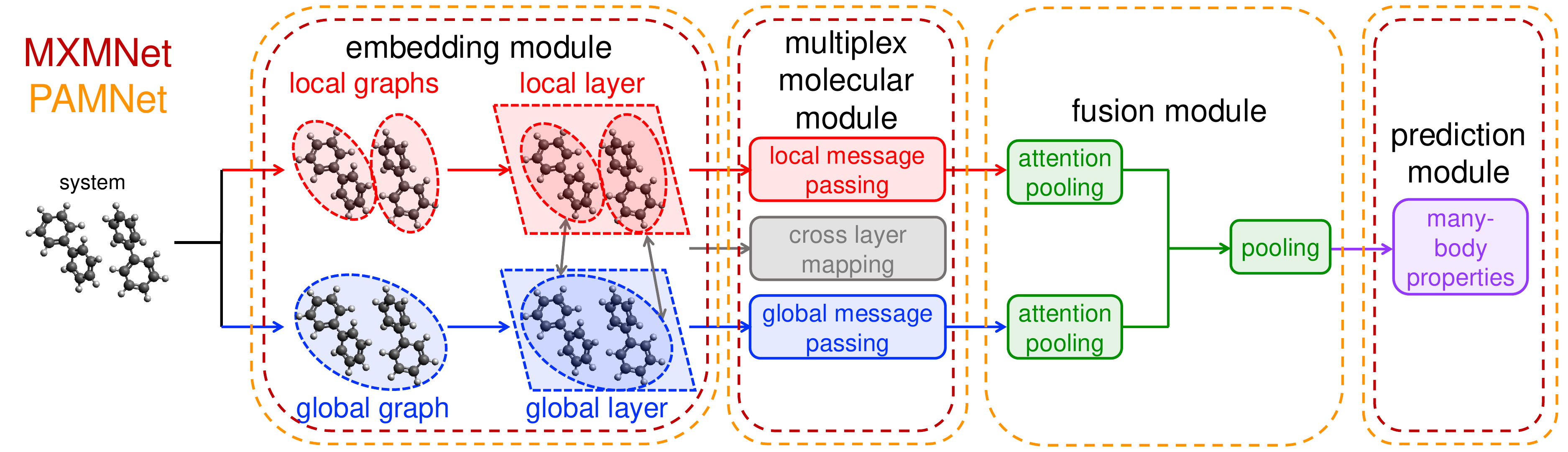}
\caption{Schematic design of MXMNet (burgundy boxes) and PAMNet (orange boxes) for a multi-fragment complex system~\cite{zhang2020molecular,zhang2023universal}.}
\label{fig:2}
\end{figure}

\paragraph{Backbone MXMNet and PAMNet Models} % as Backbone Fragment-Based Graph Neural Networks}
%\zhou{Siqi: My comment is similar to the related work section. When we discussed MXMNet and PAMNet, make sure we mention about the subgraphs. }
%\xianqi{I think we need a transitional sentence here. How about this one? “
We employed MXMNet~\cite{zhang2020molecular} and PAMNet~\cite{zhang2023universal} as our backbone FBGNN models.
Building on the general GNN framework, 
%fragment-based %GNNs %models 
%such as 
MXMNet %~\cite{zhang2020molecular} 
and PAMNet %~\cite{zhang2023universal} 
leveraged multiplex global--local architectures to align with hierarchic chemical systems. % of a global graph and many local graphs (subgraphs).
They represented the entire material as the global graph ($\mathcal{G}_\text{g}$) and every single building block %single molecule, monomer, or other fragment 
as a local graph ($\mathcal{G}_\text{l}$, or subgraph).
They also represented short-range interatomic interactions as local edges ($\mathcal{E}_\text{l}$) within a local graph, and long-range interfragment interactions as global edges ($\mathcal{E}_\text{g}$) between local graphs. 
%molecular graphs and subgraphs 
%so as to further refine the representations of short-range interatomic interactions and long-range interfragment interactions. %and non-local atomic interactions.
%”(Please help me to check the specific information again.)}
%In the present study, we 
%Here we employed both methods
%Our pioneering study employs 
%models of 
%MXMNet~\cite{zhang2020molecular} and PAMNet~\cite{zhang2023universal} %developed by Xie and coworkers 
%as our backbone FBGNN models. %subgraph-based GNNs and extended their operation to molecular graphs. 
%\zhou{Siqi: Don't they use molecular graphs?} 
%\siqi{MXMNet and PAMNet incorporate molecular graphs by representing molecules as graphs, while subgraphs are utilized to capture specific interactions or features within the larger molecular graph, allowing the models to focus on relevant local and non-local interactions for enhanced representation learning.}
Both models applied a %Following both models, we applied a 
two-layer multiplex graph % to represent our hierarchic benchmark systems 
($\mathcal{G} = {\mathcal{G}_\text{g},\mathcal{G}_\text{l}}$).
{The global graph viewed the entire system as a network of pre-defined fragments (%subgraphs or 
local graphs) and many-fragment interactions (global edges), symbolized as $\mathcal{G}_\text{g} = (\mathcal{G}_\text{l}, \mathcal{E}_\text{g})$. %~\cite{zhang2020molecular,zhang2023universal}.
Each local graph, % ($\mathcal{G}_\text{l}$), 
containing a single fragment, viewed this fragment as a network of atoms [nodes ($\mathcal{V}$)] and chemical bonds (local edges), symbolized as $\mathcal{G}_\text{l} = (\mathcal{V}, \mathcal{E}_\text{l})$}. %~\cite{jiang2021could}.} 
%This model introduces two-layer multiplex graph ($G = {G_{global},G_{local}}$) representation of  molecular systems, 
The cross mapping modules and the message passing algorithms%developed by Gilmer, Dahl and coworkers
~\cite{gilmer2017neural} integrated all information from the global and %layer and the 
local layers (Figure \ref{fig:2}). %~\cite{zhang2020molecular,zhang2023universal}. %\xianqi{We have two "and" here.}
%These treatment \xianqi{These processing?} 
This scheme allowed us to implement the geometric information about a chemical system directly into the global and local %graph and subgraph 
graph representations for downstream tasks and %, which allows us to
to
%leverages molecular structures represented by geometric information directly for quantum property predictions, 
%Local graphs represent the single-body fragment features, the global graph represents the many-body interaction features. And 
%With the cross mapping modules, we can use the properties together and have a more integrated property of the many-fragment system (Figure \ref{fig:2}). 
%\zhou{I modified your Figure 3 (originally Figure 2b) based on what you sent me earlier. The new document is Figure2-new.pptx. However, we did not label embedding, MXM, and perdition modules in the original version. I tried to modify it based on your paragraph. Could you please modify the pptx file in case I made any mistake and convert to pdf again? Thank you!}
%\xianqi{Transitional sentence: By integrating this geometric information into the overall architecture, we were able to 
seamlessly connect to the %next stage of the
next-stage model design to %, where %the 
%MXMNet and %the 
%PAMNet %framework 
further refine and process these models.

The architecture of MXMNet included three modules %an embedding module, a multiplex molecular (MXM) module, and a prediction module 
({Figure \ref{fig:2} in burgundy boxes}). 
The embedding module converted the %atomic charges and atomistic geometries as the 
Coulomb matrix~\cite{PhysRevLett.108.058301}, which contained atomic charges and interatomic distances, %as the feature of the material
%\textcolor{red}{atomic numbers (Z) and atomistic geometries} 
into a trainable embedding vector as %, which is 
the initial node feature for the multiplex graph. 
The multiplex molecular (MXM) module incorporated the local message passing mechanism, the global message passing mechanism~\cite{gilmer2017neural},  and %layer and global layer message passing mechanism, and 
the cross layer mapping for interactively updating node embeddings and local graph representation, and was the foundational component in the multiplex molecular graphs.
%The architecture foundation of both subGNN-MBE approaches, namely MXMNet-MBE and PAMNet-MBE, lies in the message passing neural network (MPNN) along with cross-layer mapping, which refines subgraph and node representations to capture both fragment-specific and system-wide properties~\cite{gilmer2017neural}. 
The prediction module leveraged the final node embeddings to predict %forecast\xianqi{I prefer use predict not forecast} 
fragment-specific and system-wide  properties in question%, such as many-body energies in the present study
~\cite{zhang2020molecular}.
%We will show that this overall %The local-global dual modeled 
An overall heterogeneous structure like this proved superior to conventional GNNs~\cite{jiang2021contrastive,wang2022survey}. % graph  %, in this approach employs local and global layers to better 
%predict model one- and many-fragment properties than conventional GNNs. %, respectively.
Building upon %the MXMNet model, 
MXMNet, PAMNet %Physics-Aware Multiplex Graph Neural Network (PAMNet) marks a significant advancement in modeling molecular structures and interactions. Its key innovation is the 
implemented an additional fusion module to %, which integrates 
integrate different types of interactions into the final prediction while ensuring the E(3)-invariance of the presentations %that the representations remain invariant under Euclidean transformations (E(3)-invariant) 
(Figure \ref{fig:2} in orange boxes). 
%This treatment \xianqi{I really not sure about this word :)}%This novol approach 
This fusion model not only enhanced the model accuracy and efficiency in capturing key features from geometric and electronic structures but also simplified the process~\cite{zhang2023universal}.
%\xianqi{Transitional sentence: 
Given the complexity of the data structure, %an effective optimization of either model is crucial. %these interactions, optimizing the models effectively is crucial. Therefore, We utilized the Adam optimizer...}
%Therefore 
we utilized the Adam optimizer%for both models 
%because it can handle sparse gradients and adaptively adjust learning rates
~\cite{kingma2014adam}%, which makes it converge faster than gradient descendent and suitable for complex chemical systems. \xianqi{Will there be a lot of CS reviewers? If so I think we'd better not make Adam so detailed.}
%\xianqi{We can keep: "we utilized the Adam optimizer 
for faster convergence and higher suitability for complex systems. %to converge models faster and make them more suitable for complex chemical systems."}
We provided the pseudo-algorithms of MXMNet and PAMNet in the context of MBE in Section \ref{sec:alg}.

\section{Experiments}

\subsection{Benchmark Systems} %set Selection and Generation}
To provide a proof-of-concept of our MXMNet-MBE and PAMNet-MBE models and assess their robustness, accuracy, and efficiency to reproduce 2B and 3B %many-fragment 
interactions, %Our first study aims to provide a proof-of-concept for the subGNN-MBE model, as well as its capability to accurately predict intermediate intermolecular interactions. 
we %need to 
established %real-life 
benchmark systems % using real molecules or materials. %, which are also ideal test cases for evaluating the accuracy and robustness of these two models.
%In the present study, we decided to 
with %We selected 
three %types of 
molecular clusters whose structures and behaviors depended on weak or moderate %intermediate 
interactions between building blocks, including pure water (\ch{H2O}) with moderate or strong hydrogen bonds, pure phenol (\ch{C6H5OH}) with weak hydrogen bonds and van der Waals interactions, and a 1:1 water--phenol (\ch{H2O:C6H5OH}) mixture showing a synergistic effect of the two interactions~\cite{gor2011matrix,zhang2016molecular}. %\xianqi{the two clusters?}
%All these systems are %focus on 
%chemical systems 
We carved all systems from condensed phases, %, whose % The 
%structures and behaviors %of these systems 
%depend on weak or intermediate interactions between building blocks, such as hydrogen bonds and van der Waals forces.
%At this phase, our research is focusing on hydrogen bonds and van der Waals forces, chosen for their essential roles in determining the structure and behavior of molecular systems. 
%Therefore we selected three molecular clusters, %with weak or intermediate intermolecular interactions like hydrogen bonds and van der Waals forces to represent the common condensed-phase systems, 
%including pure water (\ch{H2O}) with moderate or strong hydrogen bonds, pure phenol (\ch{C6H5OH}) with weak hydrogen bonds and $\pi-\pi$ stacking interactions, and a 1:1 water--phenol (\ch{H2O:C6H5OH}) mixture showing synergistic effect of the two%. These selections were made with the intention to represent a broad range of common condensed-phase interactions — water clusters for strong hydrogen bonding networks, phenol clusters for the combined effects of $\pi-\pi$ interactions and hydrogen bonding, and water-phenol mixtures for synergistic effects of weak hydrogen bonding and van der Waals interactions
%~\cite{gor2011matrix,zhang2016molecular}. 
%In these systems, we 
and designed every single water or phenol molecule as a single fragment, and collected all possible dimers and trimers for the evaluation of 2B and 3B energies in Equation (\ref{eq:mbe}). % in truncated MBE. % and truncated the MBE expression up to 3B terms. 
We calculated 1B %(one-fragment) 
energies %from Equation (\ref{eq:mbe}) 
using DFT or MP2 and predicted 2B %(two-fragment) 
and 3B %(three-fragment) 
energies using MXMNet and PAMNet models.

\subsection{Molecular Dynamics and Quantum Mechanical Calculations.}
\begin{table}[h!]
\centering
\caption{Summary of Molecular Dynamics Details}
\label{tab:md}
\begin{tabular}{c|c|c|c|c|c|c|c}
\hline
Data Set & Molecules & $T$ (K) & Dimers & Trimers & $t_\text{em}$ (ns) & $t_\text{eq}$ (ns) & $t_\text{pr}$ (ns) \\
\hline
\ch{H2O} & 67 & 370 & 48,643 & 1,053,911 & 0.5 & 0.1 & 2.0 \\
\hline
\ch{C6H5OH} & 10 & 360 & 45,045 & 120,120 & 0.5 & 0.1 & 1.0 \\
\hline
\ch{H2O:C6H5OH} & 10:10 & 694 & 190,000 & 228,000 & 0.5 & 0.1 & 0.1 \\
\hline
\end{tabular}
\end{table}

%\zhou{Siqi: If you feel this paragraph is too long, you may create a table for it.}
To sample an ergodic and sufficient data set %with sufficient for each of the three 
for each benchmark system, we included high-energy points from its FD-PES. %and sample sufficient data points for their FD-PES's, we %that reflects these interactions, we 
We employed %molecular dynamics (MD) 
MD simulations at the canonical ensemble (constant $NVT$) using GROMACS~\cite{abraham2015gromacs} at a doubled ($2\times$) density, and %ies of these systems,.
%In order to 
%To sample high-energy configurations, we doubled ($2\times$) the normal densityies of these systems,
%and 
generated their initial configurations using PACKMOL~\cite{martinez2009packmol}.
%so that we set \textcolor{red}{67 water molecules, 10 phenol molecules, and a combination of 10 water molecules and 10 phenol molecules} in the cubic box of \textcolor{red}{$1000\AA^3$ for each system, respectively,} and generated their initial configurations using PACKMOL~\cite{martinez2009packmol}.
We also set temperatures close to or above the boiling point % to \textcolor{red}{370K for water, XXX, and XXX K to XXXXXXXX and 
to ensure a faster equilibration.
%The computational details of the study involved a two-stage process initiated by Molecular Dynamics (MD) simulations utilizing GROMACS~\cite{abraham2015gromacs} initiating with 
For each simulation we performed energy minimization ($t_\text{em}$), %$NVT$-
equilibration ($t_\text{eq}$), and %$NVT$-
production ($t_\text{pr}$), and collected a large number of snapshots. 
We summarized all the details of our MD simulations in Table \ref{tab:md}.
%via the steepest descent method followed by 
%and 2-ns equilibration before we started the production runs for another \textcolor{red}{2 ns}. %in the NVT ensemble. We doubled the system's density, which achieve equilibrium easier at higher temperatures, to intensify intermolecular interactions~\cite{martinez2009packmol}. Hence, simulations were run at 694 K. The geometries of fragments and clusters were derived from these MD simulations.
%We collected \textcolor{red}{2001, 1001, and 2001 snapshots for water, ohenol, and mixture, respectively.}
From each snapshot, % with $N$ molecules, 
we collected %carved out 
all %$N(N-1)/2$ 
dimers and trimers %to study 2B energies and %$N(N-1)(N-2)/6$ 
%all trimers to study 3B energies 
%In this way, we collected a 
to create a data set with a broad representation of geometric configurations and interfragment interactions. % for the data sets of 2B and 3B energies for all three benchmark systems.
For each data set, we randomly split them into an {80:5:15 ratio for training, validation, and test sets}. % for training and testing sets across selected clusters. We generated several timestamps for each molecular system to ensure a broad representation of potential geometrical configurations and interaction states.
%\subsection{Quantum Mechanical Calculations}
To establish the training set and calibrate the {validation and test sets}, we calculated all monomer, dimer, and trimer energies using QM methods, which were MP2~\cite{PhysRev.46.618,head1988mp2} with the aug-cc-pVDZ basis set~\cite{10.1063/1.456153,kendall1992electron} for water clusters and DFT with the $\omega$B97X-D3 exchange--correlation functional~\cite{lin2013long} and the 6-311+G(d,p) basis set for phenol-involving clusters~\cite{krishnan1980self}, all in {Q-Chem 6.2}~\cite{epifanovsky2021software}. %\siqi{The initial calculations were performed using Q-Chem 5.4. However, we later redid the calculations for all pure water samples and some phenol samples using the newer Q-Chem 6.2 software. This was done because the results these calculations had very bad training results or showed values that were too different compared to other calculations.}
%Quantum mechanical modeling of fragments were performed using DFT in the Q-Chem package~\cite{epifanovsky2021software}. Single-point calculations were carried out for each fragments using MP2~\cite{head1988mp2} along with the aug-cc-pVDZ basis set~\cite{10.1063/1.456153,kendall1992electron} for the $H_{2}O$ dataset and using $\omega$B97X-D3 exchange\text{-}correlation functional~\cite{lin2013long} along with the 6-311+G(d,p) basis set~\cite{krishnan1980self} for the $PhOH$ and $H_{2}O/PhOH$ mixture datasets. 
%To ensure precision %of these calculations, 
%we set the convergence criterion to $10^{-8}$. %for the self-consistent field (SCF) calculations was set to  a threshold of $10^{-11}$. 
Following the QM %MP2 or DFT 
calculations, we calculateed 2B and 3B energies ($E^\text{2B}_{ij}$ and $E^\text{3B}_{ijk}$)  as outlined in Equations (\ref{eq:2b}) and (\ref{eq:3b}).

\subsection{Inputs and Outputs}
%{We implemented a fragment-based approach, where the molecular structures were decomposed into definedchemically meaningful fragments (such as functional groups, rings, and chains). 
The input structure of our %selected 
data sets was comprised of atom types (\textit{e.g.}, \ch{O}, \ch{H}, \ch{C}), 3D atomic coordinates %of each atom 
%in the molecular fragments 
to capture spatial relationships, and calculated 1B, 2B and 3B energies for all possible monomers, dimers, and trimers.
The output structure was comprised of FBGNN-predicted 2B and 3B energies for these configurations.

%\zhou{The following sentence was moved from Application.}
%Water systems, characterized by their ubiquitous presence and complex hydrogen-bonding networks, serve as an ideal test case for evaluating the robustness of predictive models.

\subsection{Hyperparameter Tuning}
%Our approach also embraces a systematic strategy for hyper-parameter optimization and rigorous validation and testing to ensure model reliability and generalizability. 
Hyperparameter tuning %optimization
was executed using the validation set for each benchmark system~\cite{bergstra2011algorithms}. %\xianqi{"Hyperparameter tuning ~\cite{bergstra2011algorithms}" This seems little weird.}
%In our study of the MXMNet-MBE and PAMNet-MBE models, we 
We identified %focused on 
{six} critical hyperparameters: the number of epochs ($N_\text{epoch}$), the number of convolutional layers ($N_\text{layer}$), the local cutoff distance ($D_\text{lc}$), the global cutoff distance ($D_\text{gc}$), the batch size ($N_\text{batch}$), and the learning rate ($k_\text{learn}$), because 
%\zhou{Siqi: I suggest using some letter symbols for these properties.} \xianqi{We may use the first alphabet to present these properties. Like batchsize-b, epoch-e, loss-l}
%These parameters were selected for their 
they demonstrated significant impact on GNN performance %and their importance 
in molecular modeling~\cite{jiang2021could}. 
%methodically executed for dynamic configuration of key parameters including GPU allocation, random seed, epochs, learning rate, and number of layers, facilitating precision tuning for diverse molecular scenarios and enhancing model precision and computational efficiency. By specifying parameters such as the global and local distance cutoffs, we tune our model to accommodate different molecular systems and interaction scales. 
%Our data partitioning strategy allocates 85\% for training, 5\% for validation, and 10\% for testing to optimize the learning process while ensuring robust model validation and generalizability, thereby achieving an effective balance between in-depth training and comprehensive performance evaluation. \xianqi{I think we can drop this sub-section, I think it would be a bit redundant if we wrote these specific information.}
%\siqi{In this study, hyperparameter tuning and model evaluation were conducted using distinct training, validation, and test datasets. 
%The training set was used to update the model's weights, with optimization handled by the Adam optimizer and a learning rate schedule. 
%Hyperparameter tuning was primarily based on the validation set, where the model's 
During this process %the hyperparameter tuning, 
we evaluated the model performance %was evaluated 
after each epoch by monitoring the validation loss and %. We 
employed an early stopping mechanism to %was employed 
%to stop training if the validation loss did not show improvement within a specified number of epochs to 
prevent overfitting. %The model was evaluated using the test set, which was reserved for assessing its performance on unseen data. This approach ensured that the evaluation metrics accurately reflected the model's generalization ability.%}
%\xianqi{If we mentioned Adam here that we need remove the previous one.}

%\paragraph{Online Repository}

\section{%Experiments and 
Results and Discussions}

\begin{figure}[!ht]
\centering
\includegraphics[width=0.80\textwidth]{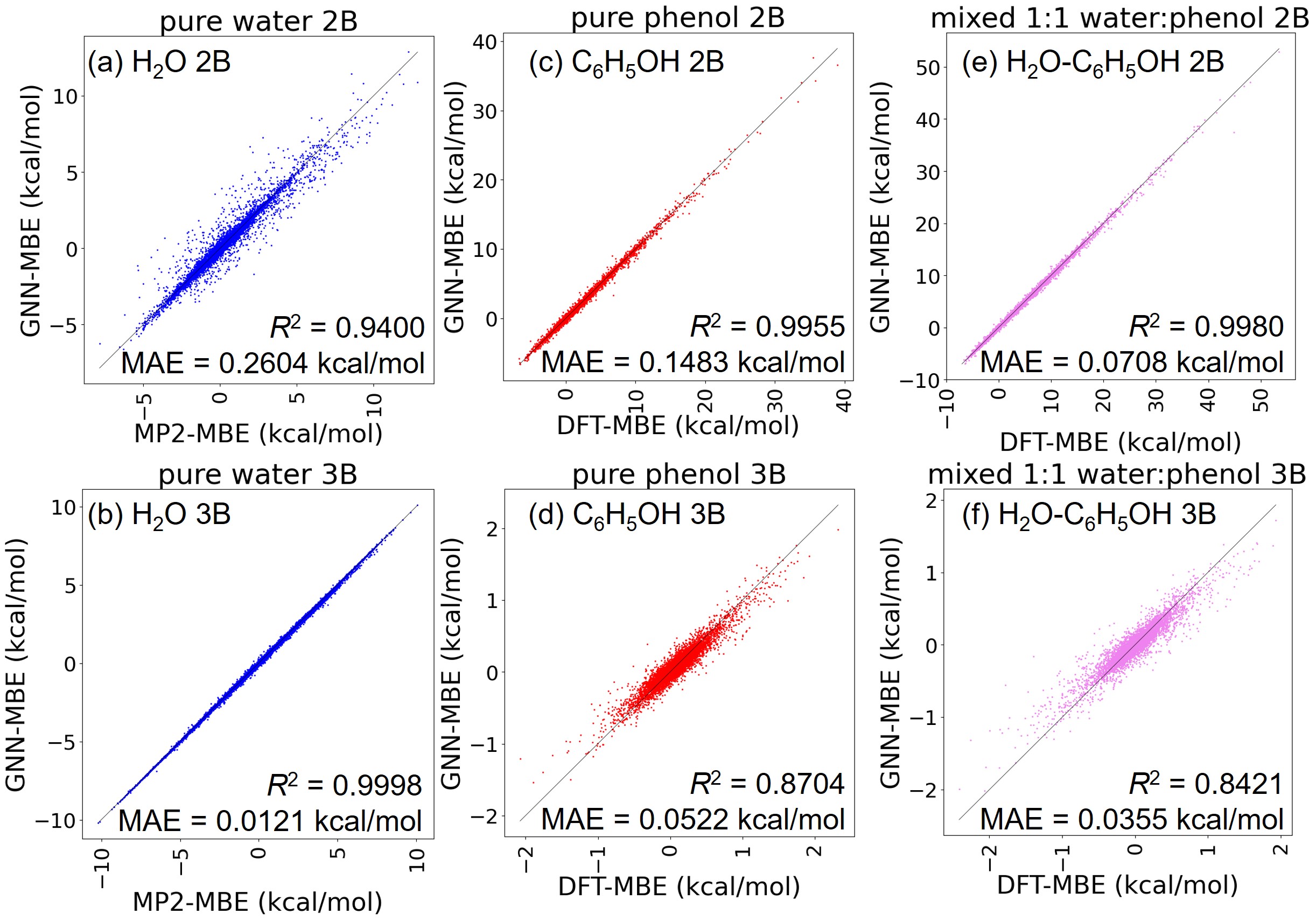}
\caption{Comparison between MXMNet-predicted and MP2/DFT-evaluated 2B and 3B energies for all three benchmark systems.} % for the clusters of pure water, pure phenol, and 1:1 water--phenol mixture.}
\label{fig:MXMNet_results}
\end{figure}

\begin{figure}[!ht]
\centering
\includegraphics[width=0.80\textwidth]{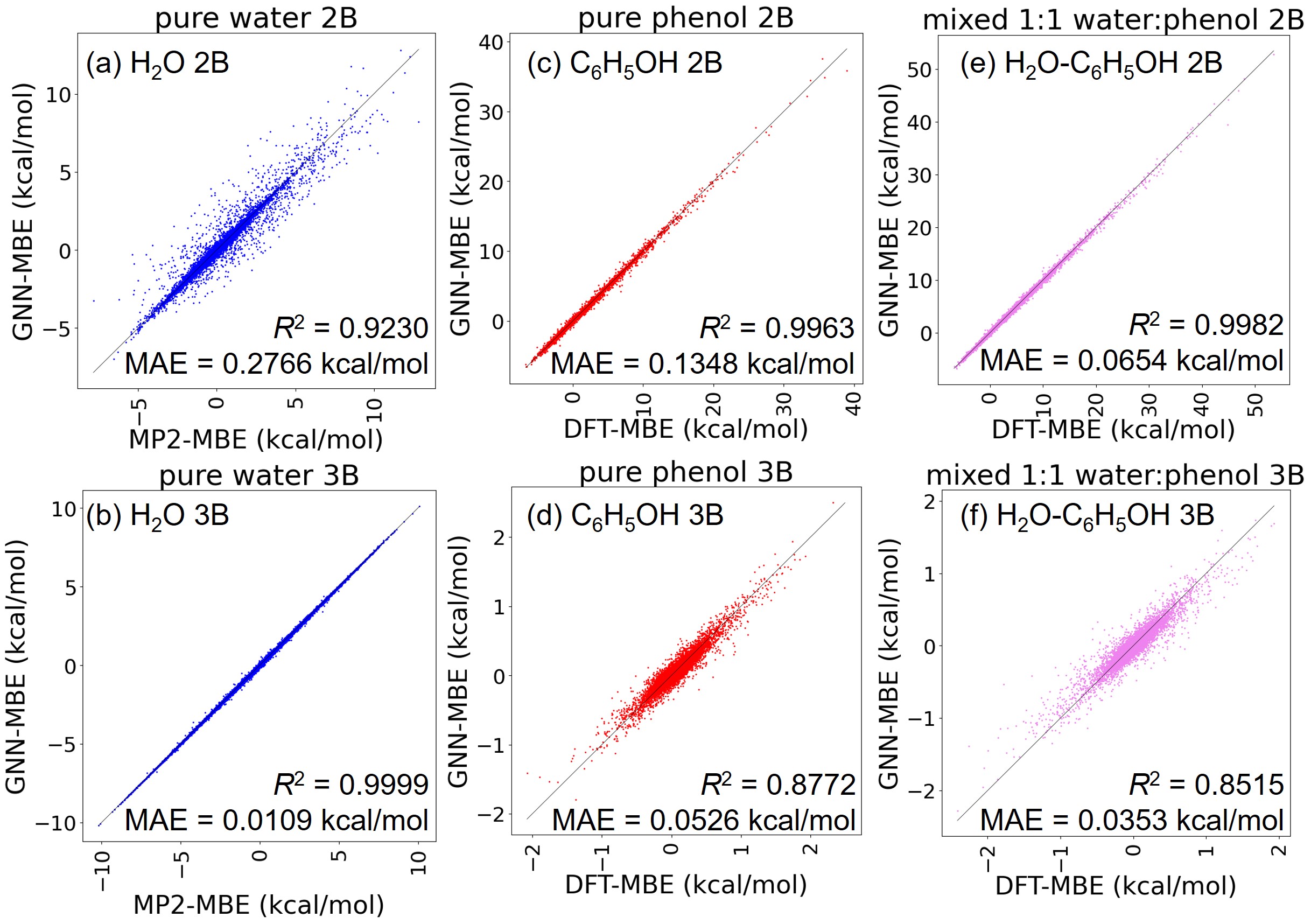}
\caption{Comparison between PAMNet-predicted and MP2/DFT-evaluated 2B and 3B energies for all three benchmark systems.} % for the clusters of pure water, pure phenol, and 1:1 water--phenol mixture.}
\label{fig:PAMNet_results}
\end{figure}

%\subsection{Performance of the MXMNet-MBE and PAMNet-MBE}
%In the present subsection, we 
To confirm the potential of FBGNN-MBE models in reproducing FD-PES for functional materials, we %We 
exhibited %the optimal results the 
their state-of-the-art performance in predicting %prediction of 
2B and 3B energies for all three benchmark systems. %, evaluated using two FBGNN-MBE approaches along with %.  of the pure water, pure phenol, and 1:1 water:phenol clusters, using for the six data sets mentioned in Section \ref{sec:data}, 
%the optimal set of hyperparameters.  
In Figures \ref{fig:MXMNet_results} and \ref{fig:PAMNet_results}, we compared 2B and 3B energies evaluated using MP2 or DFT with their counterparts predicted by MXMNet-MBE and PAMNet-MBE.
We also %report scattered comparison for all data sets , and 
summarized their values of $R$-squared coefficient ($R^2$), mean signed errors (MSE), MAE, and average CPU/GPU times for first-principles (FP) and GNN treatments %$\braket{t_\text{save}}$ (the average computational cost reduction) %, and provided the average 
in Table \ref{tab:comparative_performance}, with %We defined all of them 
their definitions in the Section \ref{sec:def}. %SM.
%\zhou{Siqi: When we write this part, we need to keep in mind that our data base has 2X(?) densities.}
%\siqi{
%MXMNet-MBE generally shows strong predictive accuracy, particularly for the water 3-body interaction, with an $R^2$ 0.9998 and an exceptionally low MAE of 0.0121 kcal/mol. 
%In comparison, PAMNet-MBE also demonstrates competitive performance, the water 3-body case with a $R^2$ of 0.9999 and an even lower MAE of 0.0109 kcal/mol, suggesting slight improvements over MXMNet in this scenario. 
%In this analysis, we used a double-density water dataset to better capture the interactions that occur at higher concentrations of water molecules. By doubling the density, increasing pressure transforms low-density water (LDW) into high-density water (HDW) by breaking hydrogen bonds between the first and second coordination shells, leading to significant structural changes and more linear hydrogen bonding configurations~\cite{soper2000structures}. This approach allows the models to account for the increased complexity and the amplified effects of hydrogen bonding, leading to more precise predictions of multi-body interactions. The MSE values being very close to 0 for all datasets indicates that the model's predictions are highly accurate, with minimal deviation from the actual values.
%}

\begin{table}[!ht]
    \centering
%    \scriptsize % or \small
    \caption{Comparative Performance of MXMNet-MBE and PAMNet-MBE Models}
    \begin{tabular}{c|c|c|r|r|r|r|r}
    \hline
        \multirow{2}{*}{Model} & \multirow{2}{*}{Dataset} & \multirow{2}{*}{$R^2$} & \multicolumn{1}{c|}{$\langle E_{\text{FP}} \rangle$} & \multicolumn{1}{c|}{MSE} & \multicolumn{1}{c|}{MAE} & \multicolumn{1}{c|}{$\langle t_{\text{FP}} \rangle$} & \multicolumn{1}{c}{$\langle t_{\text{GNN}} \rangle$}\\
        \cline{4-8}
        & & & \multicolumn{3}{c|}{kcal/mol} & \multicolumn{2}{c}{s} \\\hline
        \multirow{6}{*}{MXMNet} & \ch{H2O} 2B & 0.9400 & $+$0.2550 & $+$0.0015 & 0.2604 & 1.29  & 0.09 \\ %\cline{2-7}
        & \ch{H2O} 3B & 0.9998 & $-$0.0008 & $+$0.0005 & 0.0121 & 2.97 & 0.08 \\ %\cline{2-7}
        & \ch{C6H5OH} 2B & 0.9955 & $+$1.1199 & $-$0.0002 & 0.1483 & 147.51 & 8.83  \\ %\cline{2-7}
        & \ch{C6H5OH} 3B & 0.8704 & $+$0.0010 & $+$0.0007 & 0.0522 & 428.17 & 3.94 \\ %\cline{2-7}
        & \ch{H2O:C6H5OH} 2B & 0.9980 & $+$0.8180 & $-$0.0023 & 0.0684 & 69.77 & 4.67 \\ %\cline{2-7}
        & \ch{H2O:C6H5OH} 3B & 0.8421 & $-$0.0088 & 0.0000 & 0.0355 & 198.77 & 3.98 \\ \hline
        \multirow{6}{*}{PAMNet} & \ch{H2O} 2B & 0.9230 & $+$0.2550 & $+$0.0021 & 0.2766 & 1.29 & 0.11 \\ %\cline{2-7}
        & \ch{H2O} 3B & 0.9999 & $-$0.0008 & $+$0.0015 & 0.0109 & 2.97 & 0.07 \\ %\cline{2-7}
        & \ch{C6H5OH} 2B & 0.9963 & $+$1.1199 & $-$0.0075 & 0.1348 & 147.51 & 8.88 \\ %\cline{2-7}
        & \ch{C6H5OH} 3B & 0.8772 & $+$0.0010 & $+$0.0007 & 0.0526 & 428.17 & 3.35 \\ %\cline{2-7} 
        & \ch{H2O:C6H5OH} 2B & 0.9982 & $+$0.8180 & $-$0.0024 & 0.0654 & 69.77 & 4.68 \\ %\cline{2-7}
        & \ch{H2O:C6H5OH} 3B & 0.8515 & $-$0.0088 & $+$0.0001 & 0.0353 & 198.77 & 3.76 \\ \hline
    \end{tabular}
    \label{tab:comparative_performance}
\end{table}

\subsection{Overall Performance Assessment} % of MXMNet-MBE and PAMNet-MBE} %Nature of Superiority}
%\zhou{Your original title is vague...}
We herein analyzed the overall performance of MXMNet-MBE and PAMNet-MBE to provide a proof-of-concept for these two models using benchmark systems.
In the tasks of predicting 2B and 3B energies, both MXMNet-MBE and PAMNet-MBE
%In many-body expansion energy tasks, both MXMNet and PAMNet 
demonstrated extremely high accuracy and efficiency, %effectiveness in computing interaction energies between molecular fragments, 
with only subtle differences between each other. %in their performance. 
%Based on Figures \ref{fig:MXMNet_results} and \ref{fig:PAMNet_results} and Table \ref{tab:comparative_performance}, % show that %The data 
%shows that 
For 2B energies, they %both MXMNet-MBE and PAMNet-MBE 
achieved %high predictive accuracy and efficiency when predicting 2B energies, with 
$R^2>0.92${, |MSE| $< 0.003$ kcal/mol}, and MAE $<0.28$ kcal/mol %, and reduced the computational cost by more than %$\braket{t_{\text{save}}}> 
%92\%  %for pure water %values above 0.94 
for pure water, and $R^2>0.99${, |MSE| $< 0.008$ kcal/mol}, and MAE $<0.15$ kcal/mol % and reduced the computational cost by more than 99.5\% %$\braket{t_{\text{save}}}> 99.5\%$ 
for pure phenol and water--phenol % interactions of phenol and 
mixture. 
Regarding 3B energies, %we can draw the same conclusion: 
%MXMNet-MBE and PAMNet-MBE 
they arrived at the best performance of $R^2>0.999${, |MSE| $< 0.002$ kcal/mol}, and MAE $<0.013$ kcal/mol %, and $\braket{t_{\text{save}}}> 97.4\%$  %values above 0.94 
for pure water, %and this data set gave the best performance for all 
and $R^2>0.84${, |MSE| $< 0.001$ kcal/mol}, and MAE $<0.06$ kcal/mol for pure phenol and water--phenol mixture. % and $\braket{t_{\text{save}}}> 99.8\%$. % for pure phenol and water--phenol % interactions of phenol and 
%mixture. 
The values of MSEs were universally negligible %(within $\pm$0.008 kcal/mol) 
compared to the typical error bars of MP2~\cite{doi:10.1021/jp211997b} and DFT~\cite{yuk2024putting}, indicating the absence of systematic errors or biases in FBGNN-MBE.
%All these \textcolor{red}{negligible MSEs show that there is no any systematic error or bias in our FBGNN-MBE models.} 
%All these 
Similarly, all MAEs fell significantly below the threshold of chemical accuracy of 1 kcal/mol.
Along with large values of $R^2$ they implied the potential of well-trained FBGNN models %can potentially replace 
to replace MP2 or DFT in generating lower order terms in MBE %The MSE values being very close to 0 for all datasets indicates that the model's predictions are highly accurate, 
with minimal deviations from the actual values. %”}
%All these significant saves in computational cost present that FBGNN-MBE can accelerate the fragment-based calculations by two to four orders of magnitude. %\xianqi{Too many "All these". %"All these \textcolor{red}{negligible MSEs show that there is no any systematic error or bias in our FBGNN-MBE models.} 
%Similarly, the MAEs lie significantly below the definition of chemical accuracy of 1 kcal/mol, indicating that FBGNN-MBE models can potentially replace MP2-MBE or DFT-MBE in generating lower order terms.
Additionally, the significant reductions in computational costs by more than 91.5\% or higher confirmed that FBGNN-MBE can accelerate the fragment-based calculations by two to four orders of magnitude.
This was because the computational complexity was decreased from $N^5M^5$ for MP2 and $N^3M^5$ for MP2-MBE to $NM^5$ for MP2-based FBGNN-MBE, or $N^3M^3$ for DFT and DFT-MBE to $NM^3$ for DFT-based FBGNN-MBE, for a system containing %was significantly reduced.
%For example, if a system contains 
$N$ fragments with $M$ basis functions each fragment. % each, the computational cost  % because
%"}

%\zhou{Siqi: I think you can use my MBE-results-old.pptx file as a template to revise Figures 4 and 5. I have also uploaded the Jupyter Notebook I revised from yours to make my talk in July. In addition, I suggested providing also mean signed errors (MSE) because we want to show they are close to zero. It will be better if we can convert the x and y axes from a.u. to kcal/mol, if it is easier. Right now hartree is too big a unit for 2B and 3B energies. You also need to change DFT to MP2 for water clusters.}

\subsection{Comparison between MXMNet-MBE and PAMNet-MBE}
%PAMNet-MBE exhibits extremely similar behavior, except that it is overall slightly smaller (Figure \ref{fig:4} and Table \ref{tab:comparative_performance}).

%\zhou{Siqi: Could you please make sure Figures 4 and 5 are correct? I am very hesitant to claim that PAMNet outperforms MXMNet.}
We herein proved the marginally stronger performance of PAMNet-MBE compared to MXMNet-MBE due to the inclusion of the fusion module. 
Behaviors %\xianqi{The performance?} 
of MXMNet-MBE and PAMNet-MBE were similar, with PAMNet-MBE slightly outperforming MXMNet-MBE in %terms of 
accuracy and slightly underperforming in %terms of 
efficiency for 2B energies. %\siqi{(in terms of $R^2$ values and computational efficiency. For instance, PAMNet-MBE yields an impressive $R^2$ of 0.9963 on the Phenol 2-body dataset compared to MXMNet-MBE’s 0.9955, while also delivering lower MAE and MSE.)} 
For example, %for the 2B energies of 
for pure phenol, %mixture 2-body dataset, 
MXMNet-MBE attained impressive $R^2=0.9955$ and MAE $=0.1483$ kcal/mol and reduced the time cost by 99.92\%, 
while PAMNet-MBE achieved %slightly outperforms MXMNet-MBE, %particularly in the same dataset, where it 
%by achieving an 
$R^2=0.9963$ and  MAE $=0.1348$ kcal/mol and reduced the time cost by 99.88\%. 
This result indicated that the implementation of the fusion module in PAMNet enhanced the capacity to capture moderate two-fragment interactions but introduced a minor increase in the computational cost. %but can slow down the model. % the indicating superior accuracy in capturing interaction energies (except for water 2-body interactions ).
%When comparing 3-body interactions, 
%Regarding 3B energies, PAMNet-MBE almost consistently edges out MXMNet-MBE. 
%In the water 3-body dataset, 
%For pure water PAMNet-MBE achieves $R^2=0.9999$ with MSE and MAE of \textcolor{red}{X.XXXX} and 0.0109 kcal/mol, respectively, and compared to MXMNet’s R$^2$ of 0.9998 and an MAE of 0.0121 kcal/mol. 
%\zhou{Siqi: I am a bit hesitant here too, because your table seems to indicate that PAMNet can also be slower than MXMNet in some cases.}
%In addition, for 
Regarding the larger and more sophisticated 3B energy data sets, PAMNet-MBE showed a higher efficiency in addition to a higher accuracy.
%Combining these two advantages, 
From these observations we concluded that PAMNet was the preferred choice as a FBGNN for our current and future FBGNN-MBE tasks. 
%In addition, PAMNet's marginally better performance is complemented by its computational efficiency, as evidenced by faster convergence times, making it a more efficient model for simulating complex molecular interactions. These results suggest that PAMNet offers slight advantages in both accuracy and computational efficiency, positioning it as the preferred choice for these tasks.

\subsection{Comparison across Data Sets}
%\zhou{Siqi: Could you please make a statistics what are the average values of all these 2B and 3B energies and provide them in the table? I think these numbers are useful because they give us a sense what the magnitude of these values are.}
We herein discussed the relationship between the natures and strengths of 2B and 3B interactions in a system and the behaviors of PAMNet-MBE and MXMNet-MBE.
{A comparison across different data sets reflected the character of these benchmark systems in addition to the accuracy of the FBGNN-MBE models.}
%For example, %A comparison between two- and 3B energies exhibits significant difference. I
%across all data sets, the \textcolor{red}{MSE and} MAE associated with 2B energies are greater than those from 3B energies in the absolute values 
%\zhou{Siqi: I will revise the sentences once I have those data.}
For example, %magnitudes of 
all MAEs of %associated with 
2B energies %from all clusters are 
were significantly higher than %their corresponding 
3B counterparts, because the magnitudes of 2B energies were usually greater than 3B energies. % in absolute values, \textcolor{red}{but are almost the same in the percentages.}
For a similar reason, due to the stronger hydrogen bonds in pure water (a few kcal/mol), its %magnitudes of 
MAEs associated with 2B energies were also considerably higher than pure phenol and water--phenol mixture which were dominated by weaker van der Waals interactions. % in absolute values \textcolor{red}{but are almost the same in the percentages.}
%This is because that hydrogen bonds play an important role in the water cluster, which can be as strong as a few kcal/mol. 
%In pure phenol and water--phenol mixture, the densities of the hydrogen bonds are lower than pure water clusters, but the number of other van der Waals interactions increase, such as dipolar and quadropolar interactions. 
%In addition, 
Moreover, for pure water %clusters, 
%For water systems, our 
both MXMNet-MBE and PAMNet-MBE exhibited a stronger performance in 3B energies than 2B energies, 
%This observation 
which was intriguing and counterintuitive 
%an intriguing performance trend, demonstrating higher accuracy in 3-body interactions compared to 2-body interactions. This seemingly counterintuitive result 
but can be attributed to several %key 
factors. 
(a) %\xianqi{Firstly} 
Water trimers %in water 
encompassed broader ranges of geometries and QM effects and a larger data set and offered %richer and 
more informative data for model training, while dimers were usually oversimplified by missing some critical interactions. 
%\zhou{I am a bit worried about the following sentence. 2X density data set seems to be more repulsive in general. We need to clean up the paradox.}
%Second,\xianqi{Secondly} %Furthermore, 
(b) 3B terms from the $2\times$-density clusters %in MBE 
incorporated {both attractive (negative) and repulsive (positive) effects, % even with even distribution for negative and positive values, 
while the repulsion dominated in dimers, which were captured by the models.} % 
%\zhou{Siqi: you wrote originally ``cooperative and anti-cooperative effects''. What do you mean?}
%\siqi{Chemically, the superior performance of predicting 3B energies for pure water clusters can be attributed to the models' ability to more accurately capture the complex, spatially extended nature of hydrogen bonding, where multiple interactions between water molecules influence each other, which is simplified in 2B systems.}
%while absent in 2-body interactions, enabling 
%This behavior allows our models to leverage their advanced architectures to identify subtle contributions. 
%Finally, 
(c) Long-range interactions and polarizations prevailed in 3B interactions, which aligned with the model strength. %The enhanced performance may also arise from the models' improved capacity to account for long-range interactions and polarization effects, which are more prominent in 3-body water systems.

\subsection{Comparison with Conventional GNN Models}
We herein demonstrated the state-of-the-art accuracy of FBGNN-MBE.
%To demonstrate the advantages of FBGNN in MBE, %subgraph-based GNNs like MXMNet and PAMNet in our FBGNN-MBE model, 
We conducted a comparative analysis against MBE approaches built upon several %other 
established GNN architectures at the frontier of computational chemistry and molecular representation learning, % as discussed in the introduction, 
including SchNet~\cite{schutt2018schnet}, %PhysNet~\cite{unke2019physnet}, SIGN~\cite{rossi2020sign}, SGCN~\cite{huang2023sgcn}, and 
DimeNet~\cite{gasteiger2020directional}
DimeNet++~\cite{gasteiger2020fast}, and ViSNet~\cite{wang2024enhancing}. 
%\zhou{Siqi: I suspect the paper you placed here is not the original paper for DimeNet++}
Using %the data sets of 
2B and 3B energies of pure water clusters, we compared their accuracy % across these models 
in terms of the values of $R^2$ %, MSE, 
and MAEs %, and $\braket{t_\text{save}}$ %and MAE between MP2-calculated values and GNN-predicted ones, as well as the save in the computational costs (
(Table \ref{tab:my_label}). 
%\zhou{Siqi: Please fill out the numbers once you have them. Also, use the \{\}ch{} command when you write up a chemical formula, rather than simple \$\$.}
%As shown in Table \ref{tab:my_label}, 
%FBGNN-MBE %models using MXMNet and PAMNet exhibits a
%exhibited state-of-the-art performance in %improvement in %the accuracy and efficiency of 
%predicting 2B and 3B energies for %the data sets of 
%water clusters. %performance metrics for both 2-body and 3-body water datasets. 
For 2B energies, MXMNet-MBE and PAMNet-MBE achieved the highest $R^2$ values of 0.9400 and 0.9230, respectively, and the lowest MAEs of 0.2604 and 0.2766 kcal/mol, respectively.
These results substantially outperformed any other model, which all showed $R^2<0.67$ (without obvious trends) and MAE $>0.86$ kcal/mol (more than three times as much). 
% as MXMNet-MBE and PAMNet-MBE. %\siqi{respectively, significantly outperforming other models which have $R^2$ values below 0.67,} and 
%Specifically, the FBGNN-MBE model achieves 
%the lowest MAEs of 0.2604 and 0.2766 kcal/mol \siqi{respectively, compared to MAE values above 0.85 kcal/mol for other models}. In addition, the computational cost XXXXXXX. % and higher correlation coefficients (R\textsuperscript{2}), underscoring 
%\siqi{
%The FBGNN-MBE models using MXMNet and PAMNet significantly outperform conventional GNN architectures in predicting two- and 3B energies for water clusters, demonstrating higher accuracy with better $R^2$ values and lower MAE values compared to other established models like SchNet, DimeNet, and DimeNet++.
The performance gap was equally pronounced for the 3B energies, where % dataset. 
%Here, 
MXMNet-MBE and PAMNet-MBE achieved nearly-perfect $R^2$ values of 0.9998 and 0.9999, respectively, with remarkably low MAE values of 0.0121 and 0.0109 kcal/mol, respectively. 
While other models like DimeNet++-MBE also performed well on 3B energies ($R^2= 0.9986$, MAE $= 0.0214$ kcal/mol), MXMNet-MBE and PAMNet-MBE still maintained a clear edge in accuracy.
These results underlined the enhanced predictive accuracy of %MXMNet and PAMNet in 
FBGNN-MBE models for both 2B and 3B interactions in water clusters characterized by moderate-strength hydrogen bonds, due to the %. The substantial improvements in $R^2$ and MAE metrics demonstrate 
the efficacy of FBGNNs %our subgraph-based approach 
in capturing the complexities of both attractive and repulsive interactions.%}
%Our these results underline the enhanced predictive accuracy and efficiency of MXMNet and PAMNet in FBGNN-MBE models.%} 
%\xianqi{Should we explain why the other four models can only get all $R^2$ around 0.65?}

\begin{table}[!ht]
    \centering
    \caption{Comparative Performance of FBGNN-MBE with Other GNN-MBE Models.}
    \begin{tabular}{c|r|r|r|r}
    \hline
         Data Set & \multicolumn{2}{c|}{\ch{H2O} 2B} & \multicolumn{2}{c}{\ch{H2O} 3B} \\\hline
       GNN Model & \multicolumn{1}{c|}{$R^2$} & \multicolumn{1}{c|}{MAE (kcal/mol)} & \multicolumn{1}{c|}{$R^2$} & \multicolumn{1}{c}{MAE (kcal/mol)} \\ \hline
       MXMNet & 0.9400 & 0.2604 & 0.9998 & 0.0121 \\ %\cline{2-4}
        PAMNet & 0.9230 & 0.2766 & 0.9999 & 0.0109 \\ %\cline{2-4}
        SchNet & 0.6491 & 0.8756 & 0.9783 & 0.0957\\ %\cline{2-4}
        DimeNet & 0.6638 & 0.8796 & 0.9958 & 0.0240 \\ %\cline{2-4}
        DimeNet++ & 0.6545 & 0.8698 & 0.9986 & 0.0214 \\ %\cline{2-4}
        ViSNet & 0.6532 & 0.8752 &  &  \\ \hline
    \end{tabular}
    \label{tab:my_label}
\end{table}

%\zhou{Siqi: If you hate this table, you can consider showing it as an error bars in a figure.}
%{

\subsection{Deeper Performance Analysis}
%MXMNet-MBE demonstrated a strong prediction capacity for all six data sets.
%In particular, %, particularly 
%for the 3B interactions of water clusters, %water 3-body interaction, with an 
%it gave exceptional accuracy with %an exceptionally large value of 
%$R^2=0.9998$, MSE $= \textcolor{red}{0.0005}$ kcal/mol and %and an exceptionally low MAE of
%MAE $=0.0121$ kcal/mol. 
%The performance of PAMNet-MBE was slightly more competitive in this scenario, with %$R^2=0.9999$, MSE $= \textcolor{red}{0.0015}$ kcal/mol and %and an exceptionally low MAE of
%MAE $=0.0109$ kcal/mol.
%In comparison, PAMNet-MBE also demonstrates competitive performance, the water 3-body case with a $R^2$ of 0.9999 and an even lower MAE of 0.0109 kcal/mol, suggesting slight improvements over MXMNet in this scenario. 
We herein conducted a deeper analysis to understand the impact of the molecular density on the character of the electronic structures.
%To deepen our analysis, we 
We reported the average 2B and 3B energies from MP2 or DFT calculations as $\braket{E_\text{FP}}$ in Table \ref{tab:comparative_performance}.
Across all benchmark systems, the values of $\braket{E_\text{FP}}$ for the 2B energies were always sizable positive values, while those for the 3B energies were either positive or negative but still negligible considering the error bars of MP2 and DFT. 
The behaviors of 2B energies were counterintuitive at first sight because we had expected attractive hydrogen bonds and van der Waals interactions, but can actually be attributed to the $2\times$ densities of these sampled clusters.
We employed such $2\times$ densities because we wanted to ensure an inclusion of %make sure \xianqi{to ensure the} 
high energy configurations %were included 
in the samples to treat the increased complexity, amplify the importance of hydrogen bonds, and promote the precise reproduction of 2B/3B interactions.
%a double-density water dataset to capture interactions at higher water molecule concentrations. 
However, this treatment also altered the physics of the interacting clusters.
For example, by increasing the density and pressure, low-density water (LDW) transitions into high-density water (HDW) through the disruption of hydrogen bonds between the first and second coordination shells, resulting in significant structural shifts and more linear hydrogen bonding configurations~\cite{soper2000structures}.
At the same time, %This approach allows the models to account for the increased complexity and the amplified effects of hydrogen bonding, leading to more precise predictions of multi-body interactions. 
%\siqi{this increase in density may lead to 
HDW illustrated a stronger {(Pauli)} repulsion between closely packed molecules to counteract the attractive hydrogen bonds. %, which could counteract the expected attractive forces, resulting 
Both factors introduced sizable positive contributions for the 2B energies~\cite{stone2013theory, hesselmann2008improved}.
The behaviors of 3B energies, though smaller in magnitude, were not significantly impacted by the high density and still reflected %.% play a subtle yet crucial role in molecular interactions, obtaining near-zero or slightly negative values that reflect 
a complex interplay of cooperative and anti-cooperative effects~
\cite{mahadevi2016cooperativity,grabowski2011covalency}.
%and both repulsive and attractive forces. 
%This observation indicates that, the overall energy contributions from three-body interactions remain minimal although the high-density configurations introduced additional complexity.}

%The behaviors \xianqi{We didn't finish yet!}

%\hui{High light the most imortant results first and ablation studies/sensivitiy anslysis goes to the end. Each evaluation has its goal clarified at the front and starts with a subsection.}

\subsection{Sensitivity Analysis} % over Varying Hyperparameters}
We herein understood and optimized the model reliability %and interpretability 
of FBGNN-MBE %models in %terms of
%reliability and interpretability %of FBGNN-MBE FBGNN-MBE approaches 
in complex tasks like predicting 2B and 3B energies~\cite{agarwal2023evaluating} to guide future design and improvement. 
We analyzed the sensitivity of FBGNN-MBE approaches over varying hyperparameters. %performed the sensitivity analysis %is essential for understanding and optimizing
%to % molecular property prediction and other complex tasks. 
%We focused on the reliability and interpretability of MXMNet and PAMNet~\cite{agarwal2023evaluating} and
%As GNNs become increasingly important in fields like bioinformatics and chemistry, the demand for reliable and interpretable models is growing~\cite{agarwal2023evaluating} (Agarwal, C., Queen, O., Lakkaraju, H., \& Zitnik, M. (2023). Evaluating explainability for graph neural networks. Scientific Data, 10(1), 144.). 
%This analysis provides 
%will provide crucial insights into how various hyperparameters affect model performance to guide future design and improvement. %, guiding researchers in designing more robust and accurate GNN architectures. 
%We analyzed the sensitivity %analysis using 
%evaluation metrics across different \textcolor{red}{configurations}, such as the 
We used %the %$R$-squared coefficient (
$R^2$ coefficients %) %, as well as the \textcolor{red}{mean signed error (MSE) and} 
and MAEs of 2B and 3B energies as evaluation metrics. %to evaluate model performance across different configurations using both MXMNet and PAMNet models.
%\zhou{Siqi: You need to be more specific about what systems and methods are used for hyperparameter tuning/test. Do you use training or validation or test? Do you plan to compare MSE or it does not matter?}
%\siqi{maybe  MAE is enough for measurement of average error.}
We presented detailed discussions about these hyperparameters in Section \ref{sec:analysis}, and summarized optimized values in Table \ref{tab:hyper}.

\begin{table}[!ht]
\centering
\caption{Summary of Optimized Hyperparameters}
\label{tab:hyper}
\begin{tabular}{c|c|c|c|c|c|c|c}
\hline
Model & Data set & $N_\text{epoch}$ & $N_\text{layer}$ & $D_\text{gc}$ (\AA) & $D_\text{lc}$ (\AA) & $N_\text{batch}$ & $k_\text{learn}$ \\
\hline
\multirow{6}{*}{MXMNet} 
& \ch{H2O} 2B & 104 & 4 & 5.0 & 1.7 & 64 & 0.0001 \\
& \ch{H2O} 3B & 43 & 4 & 5.0 & 1.7 & 64 & 0.0001 \\
& \ch{C6H5OH} 2B & 154 & 4 & 15.0 & {3.0} & 64 & 0.0001 \\
& \ch{C6H5OH} 3B & 435 & 6 & 15.0 & {5.0} & 64 & 0.0001 \\
& \ch{H2O:C6H5OH} 2B & 494 & 4 & 10.0 & 5.0 & 64 & 0.0001 \\
& \ch{H2O:C6H5OH} 3B & 240 & 4 & 15.0 & 3.0 & 64 & 0.0001 \\
\hline
\multirow{6}{*}{PAMNet} 
& \ch{H2O} 2B & 207 & 2 & 5.0 & 1.7 & 64 & 0.0001 \\
& \ch{H2O} 3B & 66 & 6 & 5.0 & 1.7 & 64 & 0.0001 \\
& \ch{C6H5OH} 2B & 256 & 4 & 15.0 & {3.0} & 64 & 0.0001 \\
& \ch{C6H5OH} 3B & 54 & 3 & 15.0 & {5.0} & 64 & 0.0001 \\
& \ch{H2O:C6H5OH} 2B & 464 & 4 & 15.0 & 3.0 & 64 & 0.0001 \\
& \ch{H2O:C6H5OH} 3B & 78 & 4 & 15.0 & 4.0 & 64 & 0.0001 \\
\hline
\end{tabular}
\end{table}

%\clearpage

%\input{6_application}

\section{Conclusions and Future Directions}
%\siqi{
%In the present study, we 
We presented FBGNN-MBE, a novel %innovative 
%hybrid 
computational framework that hybridized FBGNNs with the MBE theory and exhibited %by employing a divide-and-conquer strategy and enhancing
%which 
enhanced robustness, accuracy, transferability, and interpretability from conventional NN- and GNN-accelerated QM models.
Our method addressed the prohibitive computational costs of pure QM or QM-MBE methods in modeling aggregate and dynamic properties of large %multi-fragment 
functional materials. % that remain prohibitive for %first-principles 
%pure QM or even QM-MBE approaches.
In contrast to existing QM-MBE, we only evaluated 1B energies from first principles %first-principles methods like MP2 or DFT, 
but generated $n$B ($n\ge2$) %2B and 3B 
energies based on structure--energy relationships trained by FBGNN. %-MBE.
Instead of conventional GNN models, we implemented fragment-based MXMNet and PAMNet formalisms as backbone GNN approaches % into MBE 
for a more intuitive alignment between model architecture and chemical hierarchy.
%This method 
Benchmarked on three clusters with different natures and strengths of intermolecular interactions, including pure water, pure phenol, and 1:1 water--phenol mixture,
% Our comprehensive evaluation across three benchmark systems reveals 
we revealed that well-trained FBGNN-MBE reached state-of-the-art chemical %within the chemical accuracy in
agreement with traditional QM-MBE models, with %$R^2>0.92$ and 
MAEs $<0.3$ kcal/mol for 2B energies and %$R^2>0.85$ and MAE 
$<0.02$ kcal/mol for 3B energies, but reduced  the computational cost by two to four orders of magnitude. %from QM-MBE.
%FBGNN-MBE also 
%Moreover, we accomplish a state-of-the art performance in comparison with MBE theories built upon conventional GNN models by reducing the MAE by two thirds and promoting $R^2$ by a half.
%outperformed traditional quantum mechanical models, achieving a minimum R² score of 0.92 for 2B energy predictions and 0.89 for 3B energy predictions. 

While our FBGNN-MBE framework was promising in both efficiency and accuracy, it exhibited at least two limitations that %, there are several limitations 
required our attentions %and highlight room for 
in future developments.
First, our training sets were created using $2\times$ molecular densities, which %While doubling molecular density helps 
allowed us to sample high-energy configurations and enhance ergodicity and diversity of the data set, but also introduced biases towards repulsive 2B interactions \cite{doi:10.1021/acs.chemrev.0c01111}. %, it also makes interpreting two-body interactions more complex, particularly in water clusters. 
Second, the present study focused on %our preliminary study focuses on systems 
molecular aggregates with weak to moderate many-fragment interactions and the current fragmentation strategy did not cleave chemical bonds, so that the model transferability remained elusive. 
Finally, the present study did not implement the predictions for potential energy gradients (forces) or excited state energies.
%so that we need to so we need to test systems with stronger interactions as well. 
%The PAMNet-MBE fusion module improves accuracy, however, it slow down the predictions of two-body energies. We still face ongoing challenges in obtaining sufficient training data for complex molecular systems. These limitations highlight key areas for future improvements.
Looking ahead, we will apply FBGNN-MBE in the place of QM and QM-MBE in producing %total %
%aim to extend this framework to predict total cluster 
%ground state energies, full-dimensional potential energy surfaces, 
aggregate and dynamic properties that require on-the-fly evaluations of a FD-PES, such as Monte Carlo (MC) and MD simulations, and will extend the present framework beyond ground state electronic energies, such as optical band gaps and vibrational frequencies. 
We will enhance the model transferability to various chemical systems by implementing transfer learning \cite{buterez2024transfer}.
We will improve the model capacity in physical interpretation by %of FBGNN-MBE, we also plan to 
implementing %the idea of 
the energy decomposition analysis (EDA) \cite{https://doi.org/10.1002/wcms.1345} so that we can quantitatively fraction the $n$B energies into different attractive and repulsive terms, such as electrostatics, Pauli repulsion, exchange--correlation, and %polarization, charge-transfer, and 
dispersion.  
These applications will enhance the model potentials %capacity and iterpretability %enhancing its applicability 
in the rational design of next-generation functional materials. % and  %These future directions underscore FBGNN-MBE's potential to 
%drive future innovations in computational materials science.%}
%By implementing EDA to our PES and training for all components, we expect to inject physical information to FBGNN-MBE and enhance the model understanding of 
%enhance the model performance of FBGNN-MBE can represent the balance between attraction and repulsion in hydrogen bonding and van der Waals interactions. 
%This decomposition aligns well with energy decomposition analysis (EDA), providing insights into the contribution of each force component to the overall potential energy. It enables our model to understand the 
%the physical behaviors of complex organic systems. % accurately. \\

\section{Acknowledgement}
Z.L. and H.G. thank the financial support provided by UMass Amherst Start-Up Funds 
%This work was supported by the 
%Start-Up Fund of and the 
and NSF-UMass ADVANCE Collaborative Research Seed Grant.
All authors thank the high-performance supercomputing resources provided by UMass/URI Unity Cluster and MIT Supercloud \cite{8547629}.
%, which enables us to advance this research and contribute to the field.

%\section*{References}

%\clearpage

%References follow the acknowledgments in the camera-ready paper. Use unnumbered first-level heading for the references. Any choice of citation style is acceptable as long as you are consistent. It is permissible to reduce the font size to \verb+small+ (9 point) when listing the references. Note that the Reference section does not count towards the page limit.
\medskip

{
\small
\printbibliography
%\bibliography{neurips_2024.bib}
%\bibliographystyle{IEEETR} %neurips_2024} %ieeetr
%[1] Alexander, J.A.\ \& Mozer, M.C.\ (1995) Template-based algorithms for connectionist rule extraction. In G.\ Tesauro, D.S.\ Touretzky and T.K.\ Leen (eds.), {\it Advances in Neural Information Processing Systems 7}, pp.\ 609--616. Cambridge, MA: MIT Press.
%[2] Bower, J.M.\ \& Beeman, D.\ (1995) {\it The Book of GENESIS: Exploring   Realistic Neural Models with the GEneral NEural SImulation System.}  New York: TELOS/Springer--Verlag.
%[3] Hasselmo, M.E., Schnell, E.\ \& Barkai, E.\ (1995) Dynamics of learning and recall at excitatory recurrent synapses and cholinergic modulation in rat hippocampal region CA3. {\it Journal of Neuroscience} \textbf{15}(7):5249-5262.
}

%%%%%%%%%%%%%%%%%%%%%%%%%%%%%%%%%%%%%%%%%%%%%%%%%%%%%%%%%%%%

\clearpage

\appendix

\section{Appendix} %Supplemental Material}

\subsection{Source Codes and Results}
\label{sec:alg}

\paragraph{Online Repository}
We uploaded all source codes and data sets associated with our FBGNN-MBE models, % for training, validation, and test procedures 
along with their results to the online repositories on Google Drive \url{https://drive.google.com/drive/u/1/folders/1ZY1EUURG4hD80MgXuxkp7Q0-aaZZ4MtI} and on GitHub \url{https://github.com/Lin-Group-at-UMass/FBGNN-MBE}.

\paragraph{Pseudo-Algorithms}
To make sure the readers can reproduce our MXMNet-MBE and PAMNet-MBE models, we provided their protocols as the following pseudo-algorithms. % that outline the processes used in MXMNet-MBE and PAMNet-MBE:

\begin{algorithm}[H]
\SetAlgoLined
\DontPrintSemicolon
\KwIn{Molecule data, including atom types, geometries (3D coordinates), and energies (1B, 2B, 3B) derived from MBE}
\KwOut{Predicted 2B and 3B energies}
\caption{MXMNet-MBE}

\textbf{Step 1:} Initialize embeddings and compute geometric features\\
$h \Rightarrow \text{initialize\_node\_embeddings}(x)$\; 
$(\text{edge\_index}_g, \text{dist}_g) \Rightarrow \text{radius}(\text{pos}, \text{cutoff}_g)$\;
$(\text{edge\_index}_l, \text{dist}_l) \Rightarrow \text{remove\_self\_loops}(\text{edge\_index})$\;
$\text{idx\_angles} \Rightarrow \text{compute\_angle\_indices}(\text{edge\_index}_l)$\;

\textbf{Step 2:} Encode geometric information\\
$\text{rbf}_g \Rightarrow \text{BesselBasis}(\text{dist}_g)$\;
$\text{rbf}_l \Rightarrow \text{BesselBasis}(\text{dist}_l)$\;
$\text{angle}_1 \Rightarrow \text{compute\_angles}(\text{pos}, \text{idx\_angles.two\_hop})$\;
$\text{angle}_2 \Rightarrow \text{compute\_angles}(\text{pos}, \text{idx\_angles.one\_hop})$\;
$\text{sbf}_1 \Rightarrow \text{SphericalBasis}(\text{dist}_l, \text{angle}_1)$\;
$\text{sbf}_2 \Rightarrow \text{SphericalBasis}(\text{dist}_l, \text{angle}_2)$\;

\textbf{Step 3:} Message passing layers\\
$\text{node\_sum} \Rightarrow 0$\;
\For{$l \Rightarrow 1$ \KwTo $n\_layers$}{
    \text{Global message passing}\\
    $h \Rightarrow \text{GlobalMP}(h, \text{rbf}_g, \text{edge\_index}_g)$\;
    \text{Local message passing}\\
    $(h, t) \Rightarrow \text{LocalMP}(h, \text{rbf}_l, \text{sbf}_1, \text{sbf}_2, \text{idx\_angles})$\;
    $\text{node\_sum} \Rightarrow \text{node\_sum} + t$\;
}

\textbf{Step 4:} Global pooling and prediction
$\text{output} \Rightarrow \text{global\_add\_pool}(\text{node\_sum}, \text{batch})$\;
\Return{output}
\end{algorithm}
\clearpage

\begin{algorithm}[H]
\SetAlgoLined
\DontPrintSemicolon
\KwIn{Molecule data, including atom types, geometries (3D coordinates), and energies (1B, 2B, 3B) derived from MBE}
\KwOut{Predicted 2B and 3B energies}
\caption{PAMNet-MBE}

\textbf{Step 1:} Initialize embeddings and compute geometric features\\
$h \Rightarrow \text{initialize\_node\_embeddings}(x)$\;
$(\text{edge\_index}_g, \text{dist}_g) \Rightarrow \text{radius}(\text{pos}, \text{cutoff}_g)$\;
$(\text{edge\_index}_l, \text{dist}_l) \Rightarrow \text{remove\_self\_loops}(\text{edge\_index})$\;
$\text{idx\_angles} \Rightarrow \text{compute\_angle\_indices}(\text{edge\_index}_l)$\;

\textbf{Step 2:} Encode geometric information\\
$\text{rbf}_g \Rightarrow \text{BesselBasis}(\text{dist}_g)$\;
$\text{rbf}_l \Rightarrow \text{BesselBasis}(\text{dist}_l)$\;
$\text{angle}_1 \Rightarrow \text{compute\_angles}(\text{pos}, \text{idx\_angles.two\_hop})$\;
$\text{angle}_2 \Rightarrow \text{compute\_angles}(\text{pos}, \text{idx\_angles.one\_hop})$\;
$\text{sbf}_1 \Rightarrow \text{SphericalBasis}(\text{dist}_l, \text{angle}_1)$\;
$\text{sbf}_2 \Rightarrow \text{SphericalBasis}(\text{dist}_l, \text{angle}_2)$\;

\textbf{Step 3:} Message passing with attention\\
$\text{out\_global} \Rightarrow []$, $\text{out\_local} \Rightarrow []$\;
$\text{att\_global} \Rightarrow []$, $\text{att\_local} \Rightarrow []$\;

\For{$l \Rightarrow 1$ \KwTo $n\_layers$}{
    \text{Global message passing}\\
    $(h, \text{out}_g, \text{att}_g) \Rightarrow \text{GlobalMP}(h, \text{rbf}_g, \text{edge\_index}_g)$\;
    Append $\text{out}_g$ to $\text{out\_global}$\;
    Append $\text{att}_g$ to $\text{att\_global}$\;
    
    \text{Local message passing}\\
    $(h, \text{out}_l, \text{att}_l) \Rightarrow \text{LocalMP}(h, \text{rbf}_l, \text{sbf}_2, \text{sbf}_1, \text{idx\_angles})$\;
    Append $\text{out}_l$ to $\text{out\_local}$\;
    Append $\text{att}_l$ to $\text{att\_local}$\;
}

\textbf{Step 4:} Feature fusion with attention\\
$\text{att\_scores} \Rightarrow \text{concat}(\text{att\_global}, \text{att\_local})$\;
$\text{att\_weights} \Rightarrow \text{softmax}(\text{LeakyReLU}(\text{att\_scores}))$\;
$\text{out} \Rightarrow \text{concat}(\text{out\_global}, \text{out\_local})$\;
$\text{out} \Rightarrow (\text{out} \cdot \text{att\_weights}).\text{sum}()$\;

\textbf{Step 5:} Final pooling and prediction\\
$\text{output} \Rightarrow \text{global\_pool}(\text{out}, \text{batch})$\;
\Return{output}
\end{algorithm}

\subsection{Fragmentation Strategies}
\label{sec:strategy}

At the current development stage of FBGNN-MBE,
%In our method, 
the fragmentation strategy was determined case-by-case and depended on the chemical properties the systems in question.
We currently focused on the total ground state energy of a system and will investigate the excited state energy soon.
%So far, 
We have so far implemented a ``top-down'' fragmentation strategy with two principles: % in our fragmentation strategy. 
(1) We maintain the smallest functionally meaningful unit. 
(2) %We do not break any functional unit. and the underlying principle of not breaking the functional units and 
We break only single bonds %covalent or ionic bonds 
or non-bonding interactions \cite{D0CP06528E}. % like hydrogen bonds or van der Waals interactions. 
%Examples include a single molecule in a molecular aggregate or molecular crystal,
%a single cation or anion in a ionic liquid or ionic crystal, a monomer or repeating unit in a polymer or copolymer, and a chromophore and side chains in an organic dye.
%chemical bonds with a bond order more than two. 

In the present proof-of-concept work we studied water aggregates, phenol aggregates, and water--phenol mixtures. In these systems
%we focused on total energies of molecular aggregates, so that each 
every single molecule was a natural fragment and an intermolecular interaction like hydrogen bond and van der Waals force plays an essential role in $n$-body energies ($n\ge 2$). %In our future studies for polymers, we plan to treat each monomer or similar repeating unit as a single fragment.
For our future studies about organic polymers, we have tentatively planned to extend the fragmentation strategy %to organic polymers, and we tentatively plan to 
by treating each monomer as a fragment, cleaving the carbon--carbon (\ch{C-C}) bonds, and considering each solvent molecule as a fragment too (if any). 
We will try our best not to break any complete functional groups like phenyl (\ch{-C6H5}) and carboxylic acid (\ch{-COOH}) or known unit like an amino acid and a DNA base.

For a new system, we will run a low-level first-principles MBE$_2$ calculation where the total energy is truncated at the two-fragment interactions to make sure the fragmentation does not introduce huge estimated error to damage the chemistry.
In this way the system will maintain the smallest functional or repeating unit such as a full $\pi$-conjugation, and our model can capture short-range and long-range interactions to simulate the real chemical systems, such as covalent bonds, ionic bonds, hydrogen bonds, London dispersions, dipole--dipole interactions, $\pi$--$\pi$ stacking, and solvation effects.

If a system does not exhibit natural or obvious fragments, such as polyacetylene and polytethlyene, we will consider %utilize 
an alternative ``bottom-up'' strategy, where we will construct fragments from individual atoms, functional groups, or monomers until some convergence is reached using a low-level first-principles MBE$_2$ calculation.
In this way we will allow fragments to grow around each unit and accommodate underlying effects from local electronic environment and molecular conformations.
%This approach helps create fragments that capture the local electronic environment and accommodate various conformations, enhancing the model's accuracy in sampling conformational space. 
%For organic molecules with multiple functional groups, this strategy allows fragments to grow around each unit, preserving the chemical environment and enabling more precise modeling of reactive sites. 

We foresaw great room for us to explore and validate our fragmentation strategies and enhance our model versatility and accuracy. However, due to the time constraints, we were unfortunately not able to perform a systematic discussion in the present study. 
We did validate our top-down fragmentation strategy using a series of water clusters $(\ch{H2O})_n$ ($n=7$, $10$, $13$, $16$, and $21$) by showing first-principles MP2-MBE$_2$ and MP2-MBE$_3$ results %for water clusters with 
in Table \ref{tab:mbe_error_analysis} in Section \ref{sec:app}.
When we treated each water molecule as a single fragment, we reached an average of 3.00\% and 0.39\% relative errors for these water clusters using MP2-MBE$_2$ and MP2-MBE$_3$, confirming a favorable choice of fragmentation strategy.
%While time constraints kept us focused on our specific systems, we see great potential in these strategies for improving our model's accuracy and broader applicability, particularly complex organic compounds.

\subsection{Sensitivity Analysis}
\label{sec:analysis}
%Optionally include supplemental material (complete proofs, additional experiments and plots) in appendix. All such materials \textbf{SHOULD be included in the main submission.}

\begin{figure}[!ht]
\centering
\includegraphics[width=\textwidth]{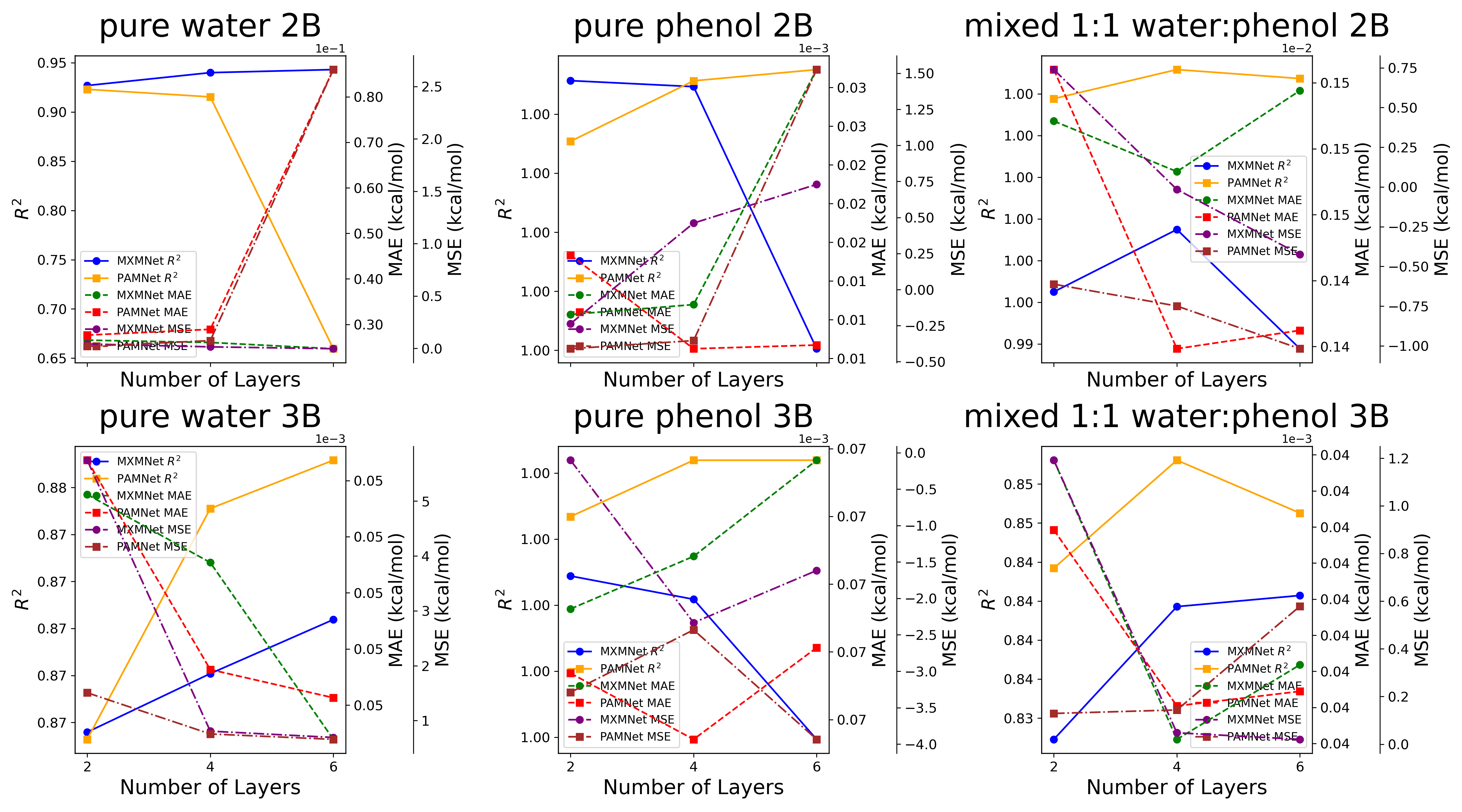}
\caption{Effects of $N_\text{layer}$ values on MXMNet-trained and PAMNet-trained values of $R^2$ and $MAE$ for all benchmark systems.} %the clusters of pure water, pure phenol, and 1:1 water--phenol mixture.}
\label{fig:number_of_layer}
\end{figure}

\paragraph{Number of Layers}
%\zhou{Siqi: I am seeing simply textural descriptions about your result. Could you please show us a table or a figure about these relationships? One picture is equal to a thousand words}
$N_\text{layer}$ had a significant impact on the capacity of a GNN model in capturing chemical features~\cite{song2021network} and its optimal value was data set-dependent. %\zhou{Siqi: Do you have a citation for this claim?} 
Our analysis found that increasing $N_\text{layer}$ from 2 to 4 %in the MXMNet model 
improved the values of $R^2$ and MAE for MXMNet-MBE and PAMNet-MBE, indicating a better fit to the training data. 
However, setting $N_\text{layer} = 6$ usually led to marginally diminishing returns
%led to diminishing returns, 
%with $N_\text{layer}=6$ %the 6-layer model
%showing 
with a slight decrease in $R^2$ and a slight increase in MAE, suggesting possible overfittings (Figure \ref{fig:number_of_layer}. 
This result supported recent findings that while a deeper GNN architecture can enhance model performance it did not always yield better generalization particularly in molecular property prediction tasks due to overfittings~\cite{Zhang2021}.
%$N_\text{layer}$ %The number of layers in a 
%GNN model has a significant effect on its ability to capture complex molecular relationships. 
%\siqi{Our analysis found that increasing $N_\text{layer}$ generally improves performance up to a point, with 4 layers often yielding the best balance between $R^2$\textcolor{red}{, MSE, and MAE}. For MXMNet-MBE, performance slightly improves or stabilizes as $N_\text{layer}$ increase from 2 to 4, indicating a better fit to the \textcolor{red}{training} data. But setting $N_\text{layer}> 4$ sometimes lead to marginal decreases in $R^2$ and increases in \textcolor{red}{MSE and MAE}, suggesting possible overfitting. PAMNet-MBE shows a more pronounced improvement in $R^2$ up to 4 layers, but adding more layers often leads to performance degradation, particularly in simpler datasets. This result supports recent findings that a deeper GNN architecture does not always yield better generalization particularly in molecular property prediction tasks~\cite{Zhang2021}. Overall, while deeper architectures can enhance model performance, they also risk overfitting, with the optimal $N_\text{layer}$ being dataset-dependent.}

\paragraph{Local and Global Cut-Off Distances}
$D_\text{lc}$ and $D_\text{gc}$ %Local and global cutoffs, which 
controled the range of interatomic, intermolecular, and interfragment interactions considered by our FBGNN-MBE. 
They were sensitive to the character of the system in question and were %scope of molecular interactions considered by the model, were 
also crucial to the model performance. 
%For MXMNet-MBE, a local cutoff of 1.7 \AA\ combined with a global cutoff of 5.0 \AA\ yielded optimal results \textcolor{red}{for the cluster of pure water}, achieving the highest $R^2$ of 0.9409 and the lowest test MAE of 0.2529 kcal/mol. 
%\siqi{
For pure water clusters, $D_\text{lc}= 1.7$ \AA\ and $D_\text{gc}= 5.0$ \AA\ provided optimal results for %particularly for 
3B interactions. 
These values confirmed that the short-range oxygen--hydrogen bond (\ch{O-H}, length 0.96\AA~\cite{brown1976geometry}) and hydrogen bonds (\ch{O-H$\cdots$ O}, length 2.5 to 4.0 \AA~\cite{scott2010new,doi:10.1021/acs.biochem.8b00217}) dominate the water clusters. %align with ranges of an oxygen--hydrogen (\ch{O-H}) bond
For example, MXMNet-MBE %, they 
reached %achieving high accuracy (MXMNet-MBE: 
$R^2 = 0.9400$ %) with low MSE and MAE values (MXMNet-MBE: 
and MAE $= 0.2604$ kcal/mol. %, MSE = 3.952*$10^{-4}$kcal/mol. ). In the case of phenol, 
For pure phenol clusters, these two values became %models required larger local cutoffs of 
3 to 5 \AA\ and %a global cutoff of 
15 \AA\ to effectively capture the long-range van der Waals interactions. %complex molecular interactions, especially in 3B systems, where these parameters improved the $R^2$ and reduced the MAE. 
For %mixed 
water-phenol mixtures, %systems, 
intermediate cutoff values balanced the short-range and long-range interactions. %, leading to strong performance in both 2B and 3B scenarios.%}
%\zhou{Siqi: Again for which system? I guess it is water? I think water and phenol may need different values. Be more specific. Also, this number seems to be ever better than your Table 1. What is going on here?}
%\textcolor{red}{These two values align perfectly with the range of a \ch{O-H} bond in XXXXX and that of a hydrogen bond in XXXXX.}
%\siqi{The selected local cutoff distance \( D_\text{lc} = 1.7 \) \AA\ for pure water clusters is well-aligned with the typical O-H bond length in water molecules, which ranges around 0.96 \AA~\cite{brown1976geometry}. This value effectively captures the intramolecular interactions within the water molecule. The global cutoff distance \( D_\text{gc} = 5 \) \AA\ encompasses the intermolecular hydrogen bonding interactions, particularly the O-O distance between water molecules, which generally lies between 2.5 \AA\ and 4.0 \AA~\cite{scott2010new}. This setup ensures accurate modeling of both short-range and long-range interactions in water clusters. }
%\zhou{Check the range of chemical bonds and hydrogen bonds and add them here into the discussion.}
This result suggested that for a system with medium interfragment interactions like hydrogen bonds, capturing moderate-range interactions provided the best balance between incorporating essential chemical information that can be missed by a smaller cut-off distance and avoiding overfitting due to a larger cut-off distance %excessively broad interaction scopes 
~\cite{Liu2024}.

\paragraph{Number of Batches}
$N_\text{batch}$ emerged %Batch size emerged 
as a key factor affecting the model performance. 
Smaller values of $N_\text{batch}$ like 64 %batch sizes, especially 64, 
consistently resulted in lower MAE values for both MXMNet-MBE and PAMNet-MBE. 
This result suggested that smaller batches facilitated more frequent weight updates and allowed the GNN models to fine-tune their parameters more effectively. 
In contrast, larger values of $N_\text{batch}$ like 256 or 512 %batch sizes (256 or 512) 
were associated with higher MAE values, likely due to less frequent updates and convergence to suboptimal solutions~\cite{ChowdhuryACM}.

\paragraph{Learning Rate}
$k_\text{learn}$ %The learning rate also had 
also impacted %a significant impact on 
the model performance. A low value of $k_\text{learn}$ like %learning rate of 
0.0001 consistently outperformed higher ones for both MXMNet-MBE and PAMNet-MBE. %models. 
For example, for the pure water data set, increasing $k_\text{learn}$ to 0.01 for PAMNet-MBE %the learning rate to 0.01 
led to a sharp decline in performance, with $R^2$ dropping to as low as {0.5981} %for MXMNet 
and a notable increase in MAE to up to {0.9468 kcal/mol}. %\siqi{Using PAMNet-MBE water 2B results here.}
This result confirmed the earlier finding that %highlights the crucial role of 
selecting an appropriate $k_\text{learn}$ can %learning rate to 
ensure stable and effective convergence during training~\cite{Zhang2021}.

\subsection{Mathematical Definition of Evaluation Metrics}
\label{sec:def}

Herein we presented the definition of all evaluation metrics used in Table \ref{tab:comparative_performance} and Figures \ref{fig:MXMNet_results} and \ref{fig:PAMNet_results}.
We set MP2/DFT evaluated 2B and 3B energies as $\{x_i\}$ and their FBGNN--predicted counterparts as $\{y_i\}$

\paragraph{$\boldsymbol{R}$-Square}

$R^2$ was the coefficient of determination between MP2/DFT and FBGNN-MBE results, defined as
\begin{equation}
    R^2 = 1- \dfrac{\displaystyle{\sum_i (y_i - x_i)^2}}{\displaystyle{\sum_i \left(y_i - \dfrac{1}{N}\sum_i y_i\right)^2}}
\end{equation}

\paragraph{Average MP2/DFT result}

The average result from first-principles (FP) MP2/DFT calculations was defined as 

\begin{equation}
\braket{E_\text{FP}}=\dfrac{1}{N}\sum_i x_i
\end{equation}
%\sum_i x_i / N$}\\

\paragraph{Mean Signed Error}
MSE was defined as the average of the signed difference between MP2/DFT and FBGNN-MBE results; as such
\begin{equation}
\text{MSE}=\dfrac{1}{N}\sum_i(y_i - x_i)
\end{equation}

\paragraph{Mean Absolute Error}
MAE was defined as the average of the unsigned difference between DFT/MP2 and FBGNN-MBE results; 
as such
\begin{equation}
\text{MAE}=\dfrac{1}{N} \sum_i |y_i - x_i|
\end{equation}

%\paragraph{Average Computational Cost Reduction}

%Average computational cost reduction is defined as \%begin{equation}
%\braket{t_{\text{save}}} = \frac{1}{N}\sum_i \frac{t (x_i)-t (y_i)}{t(x_i)}
%\end{equation}

\subsection{Future Applications}

\label{sec:app}
\paragraph{Real-Life Systems}
%\subsection{Total Energies of Water Clusters}

The ultimate goal of our development of FBGNN-MBE 
%The ultimate goal of the present study is to develop a subGNN-MBE 
formalisms is to replace QM or QM-MBE methods when an on-the-fly evaluation of FD-PBE is needed for any complex many-fragment chemical systems.
%In addition to providing a proof-of-concept for subGNN-MBE by showing its accuracy in two- and 3B energies, it 
Therefore, it is also worthwhile to validate their capacity in the real-life application scenarios %where FD-PBE is needed 
through a systematic assessment of their accuracy, efficiency, and reliability.
%\textcolor{red}{In the present section, we will evaluate the performance of two versions of subGNN-MBE in terms of their predictive power in the total energies, reduced-dimensional potential energy surfaces, and molecular dynamics trajectories of water clusters.
%In addition, herein we focused on systems with normal densities.}
%These comparative analyses will enable us to assess the accuracy and reliability of our subGNN-MBE models and determine their applicability in practical situations. 
%\zhou{Siqi: I understand that we may not finish everything, so we are always ready to cut short the paper.}
%\paragraph{Total Energies of Water Clusters}
%In the present subsection, %In this study, 
%we aimed to 
In the present study in progress, we plan to evaluate the behaviors of MXMNet-MBE and PAMNet-MBE on normal-density water clusters in reproducing three physical properties that can also be generated using MP2 and MP2-PBE, including
the total energies of water clusters [\ch{(H2O)_{\textit{n}}}], the one-dimensional (1D) projection of the FD-PES, and the trajectories of MD simulations.
These tasks require us to retrain the MXMNet-MBE and PAMNet-MBE using a mixed data set of normal and $2\times$ density clusters.
%Herein we evaluate the capacity of MXMNet-MBE and PAMNet-MBE %two versions of subGNN-MBE 
%in the %capabilities of our models in accurately predicting the 
%total energies of different %entire 
%water clusters %with different sizes 
%[\ch{(H2O)_{\textit{n}}}]. %, with the ultimate goal of applying these models to real-world industrial applications. 
We will select a series of sizes [\ch{(H2O)_{\textit{n}}}] where %a 
%water 
%The size 
the size %will 
grows from a small isolated molecular complex ($n=7$) to an actual droplet that solvates the central molecule equally to one in the liquid phase ($n=21$)~\cite{D0SC05785A}. 

In our preliminary study, we select random water complexes with $n=7$, $10$, $13$, $16$, and $21$ molecules and validate the need of performing MBE even at the MP2 level.
Here MP2-MBE$_m$ represents a MP2-based first-principles MBE truncated at the $m$-body terms, and MXMNet-MBE$_m$ and PAMNet-MBE$_m$ represents FGBNN-based MBE  truncated at the $m$-body terms as discussed in the present study.
We summarized the absolute and relative errors for MP2-MBE$_m$ ($m=1,2,3$) using these five random clusters (Table \ref{tab:mbe_error_analysis}), and estimated the upper limits for absolute and relative errors for MXMNet-MBE$_m$ and PAMNet-MBE$_m$ ($m=2,3$) by adding the errors of MP2-MBE$_m$ from Table \ref{tab:mbe_error_analysis} to the MAEs of FBGNN-MBE-generated 2B and/or 3B calculations accumulated from Table \ref{tab:comparative_performance} (Table \ref{tab:gnn_error_analysis}).
We extracted the total CPU times needed for first-principles MP2 (without MBE) using these clusters (Table \ref{tab:mbe_time_analysis}), and estimated the total CPU/GPU times needed for MXMNet-MBE$_m$ and PAMNet-MBE$_m$ ($m=2,3$) by adding the accumulated times needed for 1B calculations and the FBGNN-MBE-based 2B and 3B generations. 
%, to examine the models' performance across different cluster configurations. 
%\zhou{Siqi: We need to figure out where the water clusters came from. Did it come from some particular data set or was it generated by Cheng-Wei?} \siqi{They are from Cheng-wei. Yili mentioned in his draft "Our dataset is sampled from molecule clusters in a snapshot. "}
%Here we selected all water clusters from \textcolor{red}{XXXXXX}.
%We generated each \ch{(H2O)_{\textit{n}}} from a random snapshot of a corresponding MD simulation at the normal density and room temperature using GROMACS~\cite{abraham2015gromacs}. 
%We evaluated each benchmark total energy using MP2~\cite{PhysRev.46.618} and MP2-PBE (also truncated at the 3B terms), %, and performed first-principles MBE calculations using MP2 and truncated at the 3B terms (MP2-MBE), 
%both {at the level of aug-cc-pvdz} using Q-Chem 6.2~\cite{epifanovsky2021software}.
%We compared these results with those generated by MXMNet-MBE and PAMNet-MBE in Table \ref{tab:gcn-mbe-results} and \textcolor{red}{Figure XXXX. We found that XXXXXXXX}. % at the same truncation. % using MXMNet and PAMNet (MXMNet-MBE and PAMNet-MBE).
%\zhou{We can also do another comparison in a figure by dividing all total energies by the number of water molecules each cluster. The present table is too wide.}

\begin{table}[!ht]
\centering
\caption{Error Analysis for First-Principles MP2-MBE Using Random Water Clusters}
\begin{tabular}{c|r|r|r|r|r|r|r}
\hline
\multirow{3}{*}{Cluster} & \multicolumn{1}{c|}{Energy} & \multicolumn{6}{c}{Error}  \\
\cline{2-8}
& \multicolumn{1}{c|}{MP2} & \multicolumn{2}{c|}{MP2-MBE$_1$} & \multicolumn{2}{c|}{MP2-MBE$_2$} & \multicolumn{2}{c}{MP2-MBE$_3$} \\
\cline{2-8}
& \multicolumn{1}{c|}{hartree} & \multicolumn{1}{c|}{kcal/mol} & \multicolumn{1}{c|}{\%}  & \multicolumn{1}{c|}{kcal/mol} & \multicolumn{1}{c|}{\%} & \multicolumn{1}{c|}{kcal/mol} & \multicolumn{1}{c}{\%} \\
\hline
$\ch{(H2O)_7}$ & $-533.92$ & $63.13$ & $0.01884$ & $12.53$ & $0.00374$ & $1.36$ & $0.00041$ \\
$\ch{(H2O)_{10}}$ & $-762.77$ & $103.19$ & $0.02156$ & $22.98$ & $0.00480$ & $2.77$ & $0.00058$ \\
$\ch{(H2O)_{13}}$ & $-991.61$ & $140.18$ & $0.02253$ & $29.01$ & $0.00466$ & $3.15$ & $0.00051$ \\
$\ch{(H2O)_{16}}$ & $-1220.45$ & $180.09$ & $0.02362$ & $38.92$ & $0.00508$ & $4.03$ & $0.00053$ \\
$\ch{(H2O)_{21}}$ & $-1601.86$ & $247.90$ & $0.02466$ & $56.20$ & $0.00559$ & $10.87$ & $0.00108$ \\
\hline
\end{tabular}
\label{tab:mbe_error_analysis}
\end{table}

\begin{table}[!ht]
\centering
\caption{Estimated Maximum Error Analysis for FBGNN-MBE$_n$ Using Random Water Clusters}
\begin{tabular}{c|r|r|r|r|r|r|r|r}
\hline
\multirow{3}{*}{Cluster} & \multicolumn{8}{c}{Estimated Maximum Error} \\
\cline{2-9}
& \multicolumn{2}{c|}{MXMNet-MBE$_2$} & \multicolumn{2}{c|}{PAMNet-MBE$_2$} & \multicolumn{2}{c|}{MXMNet-MBE$_3$} & \multicolumn{2}{c}{PAMNet-MBE$_3$} \\
\cline{2-9}
& \multicolumn{1}{c|}{kcal/mol} & \multicolumn{1}{c|}{\%} & \multicolumn{1}{c|}{kcal/mol} & \multicolumn{1}{c|}{\%} & \multicolumn{1}{c|}{kcal/mol} & \multicolumn{1}{c|}{\%} & \multicolumn{1}{c|}{kcal/mol} & \multicolumn{1}{c}{\%} \\
\hline
$\ch{(H2O)_7}$ & $18.00$ & $0.00537$ & $18.38$ & $0.00549$ & $8.10$ & $0.00242$ & $8.31$ & $0.00248$ \\
$\ch{(H2O)_{10}}$ & $34.70$ & $0.00725$ & $35.43$ & $0.00740$ & $18.86$ & $0.00394$ & $19.16$ & $0.00400$ \\
$\ch{(H2O)_{13}}$ & $49.32$ & $0.00793$ & $50.58$ & $0.00813$ & $33.84$ & $0.00543$ & $34.08$ & $0.00548$ \\
$\ch{(H2O)_{16}}$ & $70.17$ & $0.00916$ & $72.11$ & $0.00942$ & $55.61$ & $0.00726$ & $55.53$ & $0.00725$ \\
$\ch{(H2O)_{21}}$ & $110.88$ & $0.01103$ & $114.29$ & $0.01137$ & $113.83$ & $0.01132$ & $112.45$ & $0.01119$\\
\hline
\end{tabular}
\label{tab:gnn_error_analysis}
\end{table}

\begin{table}[!ht]
\centering
\caption{Time Analysis for MP2-MBE$_n$ and Estimated Time Analysis for Random Water Clusters}
\begin{tabular}{c|r|r|r}
\hline
\multirow{3}{*}{Cluster} & \multicolumn{1}{c|}{MP2} & MXMNet-MBE$_3$ & PAMNet-MBE$_3$\\
\cline{2-4}
& \multicolumn{1}{c|}{CPU Time} & \multicolumn{2}{c}{Estimated CPU/GPU Time} \\
\cline{2-4}
& \multicolumn{1}{c|}{s} & \multicolumn{1}{c|}{s} & \multicolumn{1}{c}{s} \\
\hline
$\ch{(H2O)_7}$ & $1562.94$ & $9.12$ & $9.05$\\
$\ch{(H2O)_{10}}$ & $18914.93$ & $18.24$ & $17.94$\\
$\ch{(H2O)_{13}}$ & $128084.39$ & $35.73$ & $34.43$ \\
$\ch{(H2O)_{16}}$ & $51249.38$ & $62.89$ & $59.69$ \\
$\ch{(H2O)_{21}}$ & $121046.76$ & $135.07$ & $125.79$ \\
\hline
\end{tabular}
\label{tab:mbe_time_analysis}
\end{table}

From these results, we can see that the truncation at the 1B terms for water clusters introduce huge relative errors up to 0.025\%. 
(This number appears small but is still too high for a chemical accuracy.)
Inclusion of the 2B and 3B terms significantly enhance the results by reducing the relative errors to up to 0.0056\% and 0.0011\%, respectively, because an intermolecular interaction, even as weak as van der Waals force or hydrogen bond, play essential roles in chemical properties.
This conclusion validates the necessity of applying MBE theory beyond the 1B terms and suggests a truncation at 3B terms or higher. 
Due to the differences between MP2-MBE and FBGNN-MBE for these water clusters, the estimated maximum errors of FBGNN-MBE are larger than MP2-MBE but still acceptable, all below 0.011\%. 
In addition, our FBGNN-MBE calculations can reduce the computational cost by at least three orders of magnitude for water clusters compared to single-shot MP2 calculations for the whole clusters.
These two pieces of findings, although made based on estimated errors and computational costs, tentatively validates the efficiency and accuracy of FBGNN-MBE methods.

\paragraph{Planned Comparison with Other Conventional GNN Models}

In the near future, we will make a more complete assessment of the model performance of FBGNN-MBE by comparing with latest advanced %will extend our analysis to include comparisons with advanced 
fragment-based NN or GNN models designed for molecular properties, such as SE3Set~\cite{wu2024se3set}, MACE~\cite{NEURIPS2022_4a36c3c5}, %and additional fragment- and subgraph-based models, including 
MGSSL~\cite{NEURIPS2021_85267d34}, MolGAT~\cite{chaka2023high}, FragGen~\cite{yu2024fraggen}, FragNet~\cite{panapitiya2024fragnet}, GraphFP~\cite{NEURIPS2023_38ec60a9}, Subgraphormer~\cite{bar2024subgraphormer}, and ESC-GNN~\cite{yan2024efficient}.

\paragraph{Utility in Molecular Dynamics}

In the present set up, we approximated the first-principles FD-PESs, or in other words, geometry-dependent electronic energies under the Born--Oppenheimer approximation at $T=0$ K, using FBGNN-MBE-generated counterparts. 
We can evaluate all relevant forces as the gradients of these potential energies \cite{annurev:/content/journals/10.1146/annurev-physchem-062422-023532,doi:10.1021/acs.chemrev.0c01111}.
In this way, we expected to replace a first-principles FD-PES with a FBGNN-MBE-generated FD-PES in \textit{ab initio} molecular dynamics (AIMD). 
Because of this equivalent substitution, the obedience or violations to some underlying rules for FGBNN-MBE-generated forces, such as energy conservation, will depend on the overall set up of the AIMD simulation. 
If the AIMD simulation is set up at constant $NVE$ (microcanonical ensemble), the energy will be conserved.
However, if AIMD simulation is set up at constant $NVT$ (canonical ensemble) or $\mu VT$ (grand canonical ensemble), the temperature remains a constant through a computational thermostat but the energy will no longer be conserved.
%In all cases, within a single time step, we believe the total energy is conserved in the trajectory integration.

%\subsection{Fragment-Based Modeling: Approach and Pseudo-Algorithms}

%\paragraph{Fragment-Based Approach}

%Our models are designed to capture the chemical hierarchies within complex molecular systems by employing a fragment-based strategy. The fragment-based approach is like breaking down a complex molecular puzzle into meaningful pieces instead of figuring out the entire structure at once. Rather than treating the entire structure as a single huge graph, MXMNet-MBE and PAMNet-MBE divide the molecule into chemically meaningful fragments (in our prelimilary study, as individual molecules in a cluster) while not breaking strong chemical bonds or disrupt functional groups. This is essential for accurately modeling intermolecular interactions like hydrogen bonds and van der Waals forces. This method is particularly powerful for understanding how multiple molecular pieces ($n\geq2$) work together, offering both computational efficiency and chemical accuracy.

%%%%%%%%%%%%%%%%%%%%%%%%%%%%%%%%%%%%%%%%%%%%%%%%%%%%%%%%%%%%

\end{document}